\documentclass{cernyrep}
\usepackage{ifthen}
\usepackage{graphicx}
\usepackage{array,tabularx}
\usepackage{amssymb}
\usepackage{amsmath}
\usepackage{xcolor}     % for colour
\usepackage{lipsum}     % for sample text
\usepackage{ntheorem}   % for theorem-like environments
\usepackage{mdframed}   % for framing
\usepackage{enumitem}
\usepackage[bookmarks, colorlinks=true, linktoc=page, pdftex, linkcolor=black, citecolor=black, urlcolor=black]{hyperref}

\theoremstyle{break}
\theoremheaderfont{\bfseries}
\newmdtheoremenv[%
linecolor=gray,leftmargin=40,%
rightmargin=40,
backgroundcolor=gray!40,%
innertopmargin=10pt,%
ntheorem]{myinterlude}{Interlude}

\newmdtheoremenv[%
linecolor=gray,leftmargin=40,%
rightmargin=40,
backgroundcolor=red!30,%
innertopmargin=10pt,%
ntheorem]{myinterlude2}{Warning}

\newcommand{\mcos}[1]{{\mathrm{cos}}(#1)}
\newcommand{\msin}[1]{{\mathrm{sin}}(#1)}

\usepackage[small,bf,nooneline]{caption2}
\renewcommand{\captionlabeldelim}{.}

\usepackage{fancyhdr}
\fancyhfoffset{4 mm}
\fancypagestyle{ARTTITLE}{% 
\fancyhf{} % clear all header and footer fields
\lhead{\small{Proceedings of the 2018 CERN--Accelerator--School course on\\
\it{Numerical Methods for Analysis, Design and Modelling of Particle Accelerators}, Thessaloniki, (Greece)}} 
\lfoot{Available online at \url{https://cas.web.cern.ch/previous-schools}}
\rfoot{\thepage\hspace*{3mm}}
 
}

\begin{document}
\title{Mathematical and Numerical Methods for Non-linear Beam Dynamics}

\author{W. Herr}

\institute{CERN, Geneva, Switzerland}
\begin{abstract}
One of the most severe limitations in particle accelerators and beam transport are non-linear effects. Techniques to study and possibly suppress some of these detrimental effects exist, the most popular are based on particle tracking and its analysis. Studies using non-linear Normal Forms are another important and efﬁcient method. All these require a correct model of the effects and the beam dynamics in the accelerator. This lecture is an introduction to the topic, shows some of the problems and presents several contemporary tools to treat them using a systematic and consistent approach.
\end{abstract}


\maketitle % this produces the title block

\thispagestyle{ARTTITLE}

\section[Introduction]{Introduction}
\label{sec-intro}
Non-linear effects in accelerator physics are important both 
during the design stage and for successful operation of accelerators.
Since both of these aspects are closely related, they will be treated
together in this overview.
Some of the most important aspects are well described by methods
established in other areas of physics and mathematics.
Given the scope of this handbook, the treatment will be focused on
the problems in accelerators used for particle physics experiments.
Although the main emphasis will be on accelerator physics issues,
some of the aspects of more general interest will be discussed.
In particular to demonstrate that in recent years a framework has
been built to handle the complex problems in a consistent form,
technically superior and conceptually simpler than the traditional
techniques.
The need to understand the stability of particle beams has substantially
contributed to the development of new techniques and is an important
source of examples which can be verified experimentally.
Unfortunately the documentation of these developments is often
poor or even unpublished, in many cases only available as lectures or
conference proceedings.     
~~\\
This article is neither rigorous nor a complete treatment of the
topic, but rather an introduction to a limited set of contemporary tools and
methods we consider useful in accelerator theory.

\subsection{Motivation}

The most reliable tools to study (i.e. description of the machine) are simulations (e.g. tracking codes).
\vskip 1mm
\begin{itemize}[itemsep=7pt]
\item[$\bullet$] Particle Tracking is a numerical solution of the (nonlinear) Initial Value Problem.
It is a "integrator" of the equation of motion and
a vast amount of tracking codes are available, together with
analysis tools 
(Examples: Lyapunov, Chirikov, chaos detection, frequency analysis, ...)
\item[$\bullet$] It is unfortunate that theoretical and computational tools exist side by side without an undertaking how they can be integrated.
\item[$\bullet$] It should be undertaken to find an approach to link simulations with theoretical analysis, would allow a better understanding of the physics in realistic machines. 
\item[$\bullet$] A particularly promising approach is based on finite maps \cite{whnum}.
\end{itemize}
%~~\\
\newpage
\subsection{Single particle dynamics}
The concepts developed here are used to describe single particle
transverse dynamics in rings, i.e. circular accelerators or storage rings.
This is not a restriction for the application of the presented tools and
methods.
In the case of linear betatron motion the theory is rather complete
and the standard treatment \cite{cs01} suffices to describe the
dynamics. In parallel with this theory the well known concepts such
as closed orbit and Twiss parameters are introduced and emerged
automatically from the Courant-Snyder formalism \cite{cs01}.
The formalism and applications are found in many textbooks (e.g. \cite{aw01, b1,chaotig}).

In many new accelerators or storage rings (e.g. LHC) the description
of the machine with a linear formalism becomes insufficient and
the linear theory must be extended to treat non-linear effects.
The stability and confinement of the particles is not given a priori and
should rather emerge from the analysis.
Non-linear effects are a main source of performance limitations in such machines.         
A reliable treatment is required and the progress in recent years allows
to evaluate the consequences.
Very useful overview and details can be found in \cite{ef01, chao01, dragt01}.

\subsection{Layout of the treatment}  
Following a summary of the sources of non-linearities in circular machine,
the basic methods to evaluate the consequences of non-linear behaviour are discussed.
Since the traditional approach has caused misconception
and the simplifications led to wrong conclusions,
more recent and contemporary tools are introduced to treat these problems. 
An attempt is made to provide the physical picture behind these tools rather than
a rigorous mathematical description and
we shall show how the new concepts are a natural extension of
the Courant-Snyder formalism to non-linear dynamics.
An extensive treatment of these tools and many examples can be found in \cite{chao01}.
In the last part we summarize the most important physical phenomena caused
by the non-linearities in an accelerator.

\section{Variables}
For what follows one should always use canonical variables !\\
~~\\           
In Cartesian coordinates: $ {{R~=~(X, P_{X}, Y, P_{Y}, Z, P_{Z}, t)}}$
~~\\           
~~\\           
If the energy is constant (i.e. $P_{Z}$~=~const.), we use:
 ${{(X, P_{X}, Y, P_{Y}, Z, t)}}$
~~\\           
~~\\           
This system is rather inconvenient, what we want is the description of the
particle in the neighbourhood of the reference orbit/trajectory:
~~\\           
\begin{equation}
 {{R_{d}~=~(X, P_{X}, Y, P_{Y}, Z, t)}}
\end{equation}
~~\\           
which are considered now the deviations from the reference and which are zero 
for a particle on the reference trajectory
~~\\           
~~\\           
{{It is very important that it is the {\underline{reference}} {{not}} the {\underline{design}}
trajectory !}}
~~\\           
(so far it is a straight line along the Z-direction)
\subsection{Trace space and phase space}
A confusion often arises about the terms Phase Space ($x, p_{x}, ...$)~~~or~~~Trace Space ($x, x', ...$) \\
~~\\           
Although sometimes heard it is {\underline{not}} laziness or ignorance to use one or the other:
\begin{itemize}[itemsep=7pt]
\item[-]Beam dynamics is strictly correct only with ($x, p_{x}, ...$), (see later chapter) but in general quantities cannot be measured easily
\item[-]Beam dynamics with ($x, x', ...$) needs special precaution, but quantities based on these coordinates are much easier to measure
\item[-]Some quantities are different (e.g. emittance)
\end{itemize}
~\\
It comes back to a remark made at the beginning, i.e. that we shall use rings for
our arguments.
In single pass machine, e.g. linac, beam lines, spectrometers, the beam is not circulating
over many turns and several hours, therefore there is no interest in stability issues.
Instead for most of these applications what counts is the coordinates and angles
at a given position (x, x', y, y'), e.g. at the end of a beam line or a small spot one an electron
microscope.
When "accelerator physicists" talk about concepts such as tune, resonances, $\beta$-functions,
equilibrium emittances etc., all these are irrelevant for single pass machine.
There is no need to study iterating systems.
In these cases the use of the trace space is fully adequate, in fact preferred because
the quantities can be measured.
In the end, the mathematical tools are very different from the ones discussed in
this article.
~\\
\subsection{Curved coordinate system}    
For a "curved" trajectory, in {\underline{general not circular}}, with a local radius of curvature $\rho(s)$
in the horizontal (X - Z~~plane), we have to transform to a new coordinate system $(x, y, s)$ (co-moving frame) with:
~~\\
%\begin{center}
\begin{equation}   
\begin{array}{ll}
X~=~&(x + \rho) \cos\left( \dfrac{s}{\rho}\right)~-~\rho~~~~\\[3pt]
Y~=~&y\\[3pt]
Z~=~&(x + \rho) \sin\left( \dfrac{s}{\rho}\right)
\end{array}
\end{equation}   
%\end{center}
The new canonical momenta become:
~~\\
\begin{equation}   
\begin{array}{ll}
p_{x}~=~&P_{X}\cos\left( \dfrac{s}{\rho}\right)~+~P_{Z}\sin\left( \dfrac{s}{\rho}\right)\\
p_{y}~=~&P_{Y}\\
p_{s}~=~&P_{Z}\left(1~+~\dfrac{x}{\rho}\right)\cos\left( \dfrac{s}{\rho}\right)~-~P_{X}\left(1~+~\dfrac{x}{\rho}\right)\sin\left( \dfrac{s}{\rho}\right)

\end{array}
\end{equation}

\section[Sources of non-linearities]{Sources of non-linearities}
\label{sec-sources}
Any object creating non-linear electromagnetic fields on the trajectory of
the beam can strongly influence the beam dynamics.
They can be generated by the environment or by the beam itself.
\subsection{Non-linear machine elements}
Non-linear elements can be introduced into the machine on purpose or
can be the result of field imperfections. Both types can have
adverse effects on the beam stability and must be taken into account.
\subsubsection{Unwanted non-linear machine elements}
The largest fraction of machine elements are either dipole or quadrupole magnets.
In the ideal case, these types of magnets have pure dipolar or quadrupolar
fields and behave approximately as linear machine elements.
Any systematic or random deviation from this linear field introduces non-linear 
fields into the machine lattice.
These effects can dominate the aperture required and limit the stable region of the
beam.
The definition of tolerances on these imperfections is an important part of any accelerator
design.
~~\\
Normally magnets are long enough that a 2-dimensional field representation is sufficient.
The components of the magnetic field can be derived from the potential
in cylindrical coordinates ($r,~\Theta$,~s~=~0) can be written as:
\begin{equation}
 B_{r}(r, \Theta) = \sum_{n=1}^{\infty} (B_{n} {\mathrm{\sin}}(n\Theta) + A_{n} {\mathrm{\cos}}(n\Theta))  \left(\frac{r}{R_{ref}}\right)^{n-1},
\label{eq:br}
\end{equation}
\begin{equation}
 B_{\Theta}(r, \Theta) = \sum\limits_{n=1}^{\infty} (B_{n} {\mathrm{\cos}}(n\Theta) - A_{n} {\mathrm{\sin}}(n\Theta))  \left(\frac{r}{R_{ref}}\right)^{n-1},
\label{eq:bt}
\end{equation}
where $R_{ref}$ is a reference radius and $B_{n}$ and $A_{n}$ are constants.
Written in Cartesian coordinates we have:
\begin{equation}
 B(z) = \sum_{n=1}^{\infty} (B_{n} + i A_{n})  \left(\frac{r}{R_{ref}}\right)^{n-1} 
\label{eq:bcart}
\end{equation}
where $z = x + iy = r e^{i\Theta}$.
The terms $n$ correspond to 2$n$-pole magnets and the $B_{n}$ and $A_{n}$ are the
normal and skew multipole coefficients.
The beam dynamics set limits on the allowed multipole components of the installed magnets.
\subsubsection{Wanted non-linear machine elements}
In most accelerators the momentum dependent focusing of the lattice (chromaticity)
needs to be corrected with sextupoles \cite{aw01, b1}.
Sextupoles introduce non-linear fields into the lattice that are larger than
the intrinsic non-linearities of the so-called linear elements (dipoles and quadrupoles).
In a strictly periodic machine the correction can be done close to the
origin and the required sextupole strengths can be kept small.
For colliding beam accelerators usually special insertions are foreseen to
host the experiments where the dispersion is kept small and the $\beta$-function
is reduced to a minimum.
The required sextupole correction is strong and can lead to a reduction of
the dynamic aperture, i.e. the region of stability of the beam.
In most accelerators the sextupoles are the dominant source of non-linearity.
To minimize this effect is an important issue in any design of an accelerator.
~~\\
Another source of non-linearities can be octupoles used to generate amplitude
dependent detuning to provide Landau damping in case of 
instabilities.
\subsection{Beam-beam effects and space charge}
A strong source of non-linearities are the fields generated by the beam itself.
They can cause significant perturbations on the same beam (space charge effects)
or on the opposing beam (beam-beam effects) in the case of a colliding beam
facility.
~~\\
As an example, for the simplest case of round beams with the line density $n$ and
the beam size $\sigma$ the field components can be written as:
\begin{equation}\label{eq:009a}
 E_{r} = -\frac{n e}{4 \pi \epsilon_{0}}~ \cdot~ \frac{\partial}{\partial r} \int\limits_{0}^{\infty} ~~~~\dfrac{{\exp}{\textstyle{(-\dfrac{r^{2}}{(2 \sigma^{2} + q)})}}}{(2 \sigma^{2} + q)} {\mathrm{d}}q,
\end{equation}
and
\begin{equation}\label{eq:009b}
 B_{\Phi} = -\frac{n e \beta c \mu_{0}}{4 \pi }~ \cdot~ \frac{\partial}{\partial r} \int\limits_{0}^{\infty} ~~\frac{{\exp}{\textstyle{(-\dfrac{r^{2}}{(2 \sigma^{2} + q)})}}}{(2 \sigma^{2} + q)} {\mathrm{d}}q.
\end{equation}
In colliding beams with high density and small beam sizes these fields are the dominating source of
non-linearities.
The full treatment of beam-beam effects is complicated due to mutual interactions
between the two beams and a self-consistent treatment is required
in the presence of all other magnets in the ring.
\section{Map based techniques}
In the standard approach to single particle dynamics in rings, the equations of motion
are introduced together with an ansatz to solve these equations. 
In the case of linear motion, this ansatz is due to Courant-Snyder \cite{cs01}.
However, this treatment must assume that the motion of
a particle in the ring is stable and confined.
For a non-linear system this is a priori not known and the attempt to find a complete
description of the particle motion must fail.

~~\\
The starting point for the treatment of the linear dynamics in synchrotrons
is based on solving a linear differential equation of the Hill type.
\begin{equation}
 \frac{d^{2} x(s)}{d s^{2}} + \underbrace{\left( a_{0} + 2\sum_{n=1}^{\infty} a_{n}\cdot \cos(2 n s)\right)}_{K(s)} x(s) = 0~.
\end{equation}
~~\\
Each element at position $s$ acts as a source of forces,
i.e. we must write for the forces $K~~\rightarrow~~K(s)$~~~which is assumed to be a periodic function, 
i.e. $K(s + C)~=~{K(s)}_{ring}$
~~\\         
~~\\         
The solution of this Boundary Value Problem must be periodic too !
~~\\         
~~\\         
It is therefore not applicable in the general case (e.g. Linacs, Beamlines, FFAG, Recirculators, ...), much better to treat it as an {\underline{Initial Value Problem.}}\\
~~\\         
In a more useful approach we do not attempt to solve such an overall equation but rather
consider the fundamental objects of an accelerators, i.e. the machine elements themselves.
These elements, e.g. magnets or other beam elements, are the basic building blocks of
the machine.
All elements have a well defined action on a particle which can be described independent
of other elements or concepts such as closed orbit or $\beta$-functions.
Mathematically, they provide a "map" from one face of a building block to the other,
i.e. a description of how the particles move inside and between elements.
In this context, a map can be anything from linear matrices to high order integration
routines.
~~\\
A map based technique is also the basis for the treatment of particle dynamics as
an Initial value Problem (IVP).
\vskip 2mm
~~\\
It follows immediately that for a linear, 1st order equation of the type
\[
\frac{d x(s)}{d s}~=~K(s)~x(s)~~~~~~~~({\mathsf{and~initial~values~at}}~~~s_{0})             
\]
the solution can always be written as: 
%\[
\begin{equation}
\begin{array}{c}
x(s)~=~a\cdot x(s_{0})~+~ b\cdot x'(s_{0}) \\
~~\\
x'(s)~=~c\cdot x(s_{0})~+~ d\cdot x'(s_{0}) \\
\end{array}
\Longrightarrow
\left( \begin{array}{c}
x  \\
~~\\
x' \\
\end{array}\right)_{s}
~~=~~
{{
{\overbrace{
\left( \begin{array}{c}
a~~~~~~b \\
~~\\
c~~~~~~d    \\
\end{array}\right)
}^{\textstyle{{\mathsf{A}}}}}
\left( \begin{array}{c}
x  \\
~~\\
x' \\
\end{array}\right)_{s_{0}}
}}
\label{eq:eigx}
\end{equation}
%\]
~~\\
where the function $K(s)$ does not have to be periodic.
Furthermore, the determinant of the matrix $A$ is always 1.
Therefore it is an advantage to use maps (matrices) for a linear systems from the start,
without trying to solve a differential equation.

The combination of all machine elements make up the ring or beam line and it is the combination
of the associated maps which is necessary for the description and analysis of the
physical phenomena in the accelerator ring or beam line.

For a circular machine the most interesting map is the one which describes the motion once around the machine,
the so-called One-Turn-Map.
It contains all necessary information on stability, existence of closed orbit, and
optical parameters.
The reader is assumed to be familiar with this concept in the case of linear beam 
dynamics (chapter 2) where all maps are matrices and the Courant-Snyder analysis 
of the corresponding one-turn-map produces the desired information such 
as e.g. closed orbit or Twiss parameters.      
~~\\
It should therefore be the goal to generalize this concept to non-linear dynamics.
The computation of a reliable one-turn-map and the analysis of its properties
will provide all relevant information.

~~\\
Given that the non-linear maps can be rather complex objects, the analysis of the
one-turn-map should be separated from the calculation of the map itself.

\section{Linear normal forms}
\subsection{Sequence of maps}

Starting from a position {{$s_{0}$}} and combining all matrices to get the matrix to position {{$s_{0} + L$}} (shown for 1D only):
\begin{equation}
\left( \begin{array}{c}
x  \\
x' \\
\end{array}\right)_{{{s_{0}~+~L}}}
=~~
\underbrace{{{{M}}_{N}}
~~\circ~~
{{{M}}_{N-1}}
~~\circ~~
...
~~\circ~~
{{{M}}_{1}}}_{{{{M}}(s_{0}, L)}}
~~\circ~~
\left( \begin{array}{c}
x  \\
x' \\
\end{array}\right)_{{{s_{0}}}}
\end{equation}
For a ring with circumference {{C}} one obtains the {{One-Turn-Matrix}} (OTM) at {{$s_{0}$}}
\begin{equation}
\left( \begin{array}{c}
x  \\
x' \\
\end{array}\right)_{{{s_{0}~+~C}}}
=~~
\underbrace{{{\left( \begin{array}{cc}
m_{11} &m_{12}  \\
m_{21} &m_{22}    \\
\end{array}\right)}}}_{{{{M}}_{OTM}}}
~~\circ~~
\left( \begin{array}{c}
x  \\
x' \\
\end{array}\right)_{{{s_{0}}}}
\end{equation}
~~\\
Without proof, the scalar product:
\begin{equation}
\left( \begin{array}{c}
x  \\
x' \\
\end{array}\right)_{{{s_{0}}}}
\cdot 
{{{{M}}_{OTM}}}
\left( \begin{array}{c}
x  \\
x' \\
\end{array}\right)_{{{s_{0}}}}
~~=~~{\mathsf{const.~~=~~J}}
\end{equation}
is a constant of the motion: invariant of the One Turn Map.\\
~~\\
With this approach we have a strong argument that the construction of the One Turn Map
is based on the properties of each element in the machine.
It is entirely independent of the purpose of the machine and their global properties.
It is not restricted to rings or in general to circular machine.
~~\\
Once the One Turn Map is constructed, it can be analysed, but this analysis does
not depend on how it was constructed.
~~\\
As a paradigm: the construction of a map (being for a circular machine or not) and
its analysis are conceptual and computational separated undertakings.
\subsection{Analysis of the One Turn Map}
The key for the analysis is that matrices can be transformed into {\bfseries{{Normal Forms.}}}
Starting with the One-Turn-Matrix,
and try to find a (invertible) transformation {{${{A}}$}} such that:
\begin{equation}
{{A}}{{{{M}}}}{{A}}^{{-1}}~=~{{{{R}}}}
~~~~~~~~~{\mathsf{(or:}}~~~~~~~~~
{{A}}^{{-1}}{{{{R}}}}{{A}}~=~{{{{M}}}}{\mathsf{)}}
\label{eq:nform1}
\end{equation}
\vskip 2mm
\begin{itemize}
\item[-] The matrix {{${{R}}$}} is:                          
\begin{itemize}
\item[-] A "Normal Form", (or at least a very simplified form of the matrix)       
\item[-] For example (most important case): {{${{R}}$}} becomes a pure rotation     
\end{itemize}
\item[-] The matrix {{${{R}}$}} describes the same dynamics as {{${{M}}$}}, but:
\begin{itemize}
\item[-] All coordinates are transformed by {{{${A}$}}}
\item[-] This transformation {{{${A}$}}} "analyses" the complexity of the motion, it contains the structure of the phase space
\end{itemize}
\end{itemize}
\begin{equation}
 {{M}} = {{A}} \circ {{{{R}}}} \circ {{A}}^{-1}
 ~~~~{\mathrm{or:}}~~~~{{{R}}} = {{A}}^{-1} \circ {{M}} \circ {{A}}
\end{equation}
The motion on an ellipse becomes a motion on a circle (i.e. a rotation): 
${{{{R}}}}$ is the "simple" part of the map~~-~~ shape is "dumped" into ${{{{A}}}}$
the matrix ${{{{A}}}}$. The rotation matrix $R$ can be obtained by the evaluation of the Eigenvectors and Eigenvalues.
~~\\
~~\\
For a rotation in 2D the eigenvalues are of course $e^{\textstyle{\pm i \Delta \mu}}$, which also implies
that the trace becomes $Tr(R)~=~2~\cos(\Delta \mu)$.
~~\\
~~\\
It is known from basic linear algebra that $~Tr(R)~~=~~Tr(M)~$ when the matrices 
are {\underline{similar}} \footnote{Note: although sometimes claimed, the equivalence of 
matrices is {\underline{not}} a sufficient condition}(in the mathematical sense !) and in the trivial case 
this can be used to test whether the original matrix $M$ describes a rotation.
~~\\
~~\\
Rewritten in the form we want for the analysis::
\begin{equation}
{{A}} =
\left( \begin{array}{cc}
\sqrt{\beta(s)}  &0 \\[8pt]
-{\textstyle{\frac{\textstyle{\alpha(s)}}{\textstyle{\sqrt{\beta(s)}}}}} &\frac{1}{\textstyle{\sqrt{\beta(s)}}}  \\[8pt]
\end{array}\right)
~~~~~{\mathsf{and}}~~~~~{{R}} =
\left( \begin{array}{cc}
\cos(\Delta\mu) &\sin(\Delta\mu)  \\[8pt]
-\sin(\Delta\mu) &\cos(\Delta\mu)    \\[8pt]
\end{array}\right)
\label{eq:eig}
\end{equation}
~~\\
Please note that the normal form analysis gives the eigenvectors (\ref{eq:eig}) without any physical
picture related to their interpretation.
The formulation using $\alpha$ and $\beta$ is due to Courant and Snyder.
Amongst other advantages it can be used to "normalise" the position $x$:
the normalised position $x_{n}$ is the "non-normalized" divided
by $\sqrt{\beta}$. The variation of the normalised position $x_{n}$ 
is then smaller than in the non-normalized case.
This is also better suited for analytical calculation, e.g.
involving perturbation theory.
~~\\
The Normal Form transformation together with this choice gives the required information:
\begin{itemize}
\item[-] $\mu_{x}$~is~the~"tune"~~~$Q_{x}\cdot 2\pi$~~ (now we can talk about {\underline{phase advance}} !)
\item[-] $\beta, \alpha, ...$ are the optical parameters and describe the ellipse 
%(tilt, stretch \& squash)
\item[-] The closed orbit (an invariant, identical coordinates after one turn !{\footnote{not to be confused with an integer resonances}}):
\item[-] ${{{M}}_{OTM}}~\circ~(x,~x')_{co}~~\equiv~~(x,~x')_{co}$
\end{itemize}
\subsection{Action-angle variables}
More appropriate for studies of beam dynamics is the use of  Action - Angle variables.        
\vskip 1mm
~~\\
Once the particles "travel" on a circle, the motion is better described by the canonical variables action ${{J_{x}}}$ and angle ${{\Psi_{x}}}$, i.e. a normalized normal form:
\vskip 3mm
\begin{center}
\includegraphics*[width= 10.0cm,height= 5.00cm,clip=]{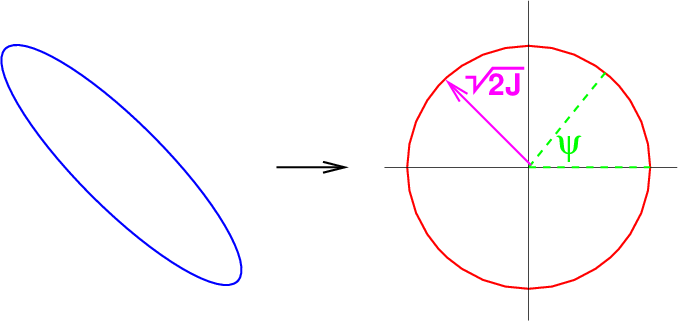}
\end{center}
with the definitions and the choice (\ref{eq:eig}) is:
\begin{equation}
\begin{array}{ll}
{\textstyle{x~= \sqrt{2 {{J_{x}}} \beta_{x}}~~\cos({{\Psi_{x}}})}}\\[8pt]
{{p_{x}~= {\textstyle{-\sqrt{\dfrac{2 {{J_{x}}}}{\beta_{x}}}}}~~( \sin({{\Psi_{x}}}) + \alpha_{x} \cos({{\Psi_{x}}}))}}\quad\\[8pt]
{{J_{x}}}~= \dfrac{1}{2} (\gamma_{x} x^{2}~+~2 \alpha_{x} x p_{x}~+~\beta_{x} p_{x}^{2})\\
\end{array}
\end{equation}
\vskip 2mm
\begin{itemize}[itemsep= 7pt]
\item[-] The angular position along the ring ${\Psi}$ becomes the independent variable !
\item[-] The trajectory of a particle is now {\underline{independent}} of the position $s$ !
\item[-] The constant radius of the circle $\sqrt{2 J}$ defines the action ${{J}}$~({{invariant~of~motion}}) \\
\end{itemize}
\subsection{Beam emittance}
A sad and dismal story in accelerator physics is the definition of the emittance.
Most foolish in this context is to relate emittance to single particles.
This is true in particular when we have a beam line which is not periodic.
In that case the Courant-Snyder parameters can be determined from the beam.
These parameters are related to the moments of the beam, e.g. the beam size
is directly related to the second order moment $<x^{2}>$.
Using the expression above for the action and angle, we can write for this
expression:
\begin{equation}
<x^{2}>~~=~~<{2 {{J_{x}}} \beta_{x}}\cdot \cos^{2}({{\Psi_{x}}})>~~=~~2\beta_{x}J_{x}\cdot <\cos^{2}({{\Psi_{x}}})>.
\label{eq:moment01}
\end{equation}
~~\\
The average of $\cos^{2}(\Psi_{x})$ can immediately be evaluated as 0.5 and defining the emittance
as:
\begin{equation}
\epsilon_{x}~~=~~<J_{x}>,
\label{eq:emitt01}
\end{equation}
and therefore the moment $<x^{2}>$ becomes:
\begin{equation}
<x^{2}>~~=~~\beta_{x}\cdot \epsilon_{x}.
\label{eq:emitt02}
\end{equation}
\newpage
Since other definitions often refer to the treatment by Courant and Snyder,
here a quote from Courant himself in \cite{cs02}:
~~\\
\begin{myinterlude}
~~\\
"The invariant $J$ is simply related to the area enclosed by the ellipse:
\begin{equation}
{\mathsf{Area~enclosed}}~~~=~~2\pi J.
\label{eq:csem01}
\end{equation}
In accelerator and storage ring terminology there is a quantity called the
$emittance$ which is closely related to this invariant. The emittance, however, is
a property of a distribution of particles, not a single particle. Consider
a Gaussian distribution in amplitudes. Then the emittance, $\epsilon$, is
given by:
\begin{equation}
(x_{rms})^{2}~~=~~\beta_{x}(s)\cdot \epsilon_{x}.
\label{eq:csem02}
\end{equation}
In terms of the action variable, $J$, this can be rewritten
\begin{equation}
\epsilon_{x}~~=~~<J>.
\label{eq:csem03}
\end{equation}
where the bracket indicates an average over the distribution in J."
~~\\
\end{myinterlude}
Using a similar procedure (details and derivation in e.g. \cite{aw01}, and to a much lesser extent in \cite{whnum}) one can determine the
moments
\begin{equation}
<p_{x}^{2}>~~=~~\gamma_{x}\cdot \epsilon_{x},
\label{eq:emitt03}
\end{equation}
\begin{equation}
{\mathsf{and}}
~~~~~~~~~<x\cdot p_{x}>~~=~~-\alpha_{x}\cdot \epsilon_{x}.~~~~~~
\label{eq:emitt04}
\end{equation}
Using these expressions, the emittance becomes readily
\begin{equation}
\epsilon_{x}~~=~~\sqrt{<x^{2}><p_{x}^{2}>~~-~~<x\cdot p_{x}>^{2}}                           
\label{eq:emitt05}
\end{equation}
Therefore, once the emittance is known, the Courant-Snyder parameters are
determined by equations (\ref{eq:emitt02}, \ref{eq:emitt03}, \ref{eq:emitt04}).
~~\\
%Other definitions based on handwaving arguments or those approximately valid only 
%in special cases, should be discarded, in particular those relying on presumed distributions, e.g. Gaussian.
\begin{myinterlude2}
~~\\
In this context it should be mentioned that frequently the invariance of (\ref{eq:csem01})
is called {\underline{the}} "Liouville theorem". 
Nothing could be more wrong as Liouville's theorem
implies much more and (\ref{eq:csem01}) is merely a rather simple consequence.
~~\\
\end{myinterlude2}

\section[Techniques and tools to evaluate and correct non-linear effects]{Techniques and tools to evaluate and correct non-linear effects}
\label{sec-techniques}
The key to a more modern approach shown in this section is to avoid the 
prejudices about the stability and other properties of the ring.
Instead, we must describe the machine in terms of the objects it consists of with all
their properties, including the non-linear elements.
The analysis will reveal the properties of the particles such as e.g. stability.
In the simplest case, the ring is made of individual machine elements such as magnets
which have an existence on their own, i.e. the interaction of a particle with a given element
is independent of the motion in the rest of the machine.
Also for the study of non-linear effects,
the description of elements should be independent of concepts such as tune, chromaticity and closed orbit.
To successfully study single particle dynamics, one must be able to describe the action
of the machine element on the particle as well as the machine element.
\subsection{Particle tracking}
The ring being a collection of maps, a particle tracking code, i.e. an integrator of
the equation of motion, is the most reliable map for the analysis of the machine.
Of course, this requires an appropriate description of the non-linear maps in the
code.
It is not the purpose of this article to describe the details of tracking codes
and the underlying philosophy, such details can be found in the literature (see e.g. \cite{ef01}).
Here we review and demonstrate the basic principles and analysis techniques.
\subsubsection{Symplecticity}
If we define a map through ${\vec{z_{2}}}~=~{{M}}_{12}(\vec{z_{1}})$
as a propagator from a location "1" to a location "2" in the ring,
we have to consider that not all possible maps are allowed.
The required property of the map is called "symplecticity" and in the simplest case
where ${{M}}_{12}$ is a matrix, the symplecticity condition can be written as:
\begin{equation}
{{M}} \Rightarrow {{M}}^{T} \cdot {S} \cdot {{M}} = {S} 
{~~~~\mathrm{where}~~~~}
{S} =
\left( \begin{array}{cccc}
0  &1  &0  &0   \\
-1 &0  &0  &0   \\
0  &0  &0  &1   \\
0  &0  &-1 &0   \\
\end{array}\right)
\end{equation}

The physical meaning of this condition is that the map is area preserving in the phase space.
The condition can easily be derived from a Hamiltonian treatment, closely related
to Liouville's theorem.

\subsection{Approximations and tools}
The concept of symplecticity is vital for the treatment of Hamiltonian systems.
This is true in particular when the stability of a system is investigated
using particle tracking.
However, in practice it is difficult to accomplish for a given exact problem.
As an example we may have the exact fields and potentials of electromagnetic
elements. For a single pass system a (slightly) non-symplectic
integrator may be sufficient, but for an iterative system the results
are meaningless.
~~\\
To track particles using the exact model\footnote{The term "model" is the description of all machine elements and the layout} may result in a non-symplectic
tracking, i.e. the underlying model is correct, but the resulting
physics is wrong.
~~\\
~~\\
It is much better to approximate the model to the extend that the tracking
is symplectic. One might compromise on the exactness of the final result,
but the correct physics is ensured.
~~\\
~~\\
As a typical example one might observe possible chaotic motion during the
tracking procedure. 
However, there is always a non-negligible probability that this interpretation 
of the results may be wrong.
To conclude that it is not a consequence of
non-symplecticity of the procedure or a numerical artifact it is necessary to
identify the {\underline{physical}} mechanism leading to this observation.
~~\\
~~\\
This may not be possible to achieve using the exact model\footnote{A non-exact model allows for small (usually irrelevant) differences such as positions, fields, length etc.} as input to a
(possibly) non-symplectic procedure.
Involving approximations to the definition of the problem should reveal the
correct physics at the expense of a (hopefully) small error.
Staying exact, the physics may be wrong.
~~\\
As a result, care must be taken to positively identify the underlying
process.
~~\\
This procedure should be based on a approximations as close as
possible to the exact problem, but allowing a symplectic evaluation.
~~\\
~~\\
\subsection{Taylor and power maps}
A non-linear element cannot be represented in the form of a linear matrix and more complicated
maps have to be introduced \cite{chaotig}.
In principle, any well behaved, non-linear function can be developed as a Taylor series.
This expansion can be truncated at the desired precision.
~~\\
Another option is the representation as Lie transformations \cite{dragt01, dragt02}.
Both are discussed in this section.
\subsubsection{Taylor maps }
A Taylor map can be written using higher order matrices and
in the case of two dimensions we have:
%\begin{equation}
%\begin{array}{lll}
% z_{1}(s_{2})= x(s_{2})  &= &~~R_{11} \cdot {{x}^{  }} ~~+ R_{12} \cdot {{x'}} ~~+ R_{13} \cdot {{y}} +  ....\\
% ~  &~ &+ T_{111} \cdot {{x^{2}}} + T_{112} \cdot {{x x'}} + T_{122} \cdot {{x'^{2}}} +\\
% ~  &~ &+ T_{113} \cdot {{x y}} + T_{114} \cdot {{x y'}} +  ....\\
% ~  &~ &+ U_{1111} \cdot {{x^{3}}} + U_{1112} \cdot {{x^{2} x'}} +  ....\\
%\end{array}
%\end{equation}
%\vskip 5mm
%and the equivalent for all other variables ...
%                 
%\vskip 2mm
\begin{equation}
z_{j}(s_{2})~=~\sum_{k=1}^{4} {{R_{jk}}} z_{k}(s_{1}) ~+\sum_{k=1}^{4}\sum_{l=1}^{4} {{T_{jkl}}} z_{k}(s_{1})z_{l}(s_{1})
\end{equation}
(where $z_{j}$, j=1,..4, stand for x, x', y, y').
Let us call the collection:~~${{{A}}_{2}}~=~({{R}},{{T}})$~~the second order map~${{{A}}_{2}}$.
Higher orders can be defined as needed, e.g. for the 3rd order map
${{{A}}_{3}}~=~(R,T,{{U}})$
we add a third order matrix:\\
\begin{equation}
+~~~\sum_{k=1}^{4}~\sum_{l=1}^{4}~\sum_{m=1}^{4}~ {{U_{jklm}}} z_{k}(s_{1})z_{l}(s_{1})z_{m}(s_{1})
\end{equation}
~~\\
Since Taylor expansions are not matrices, to provide a symplectic map,
it is the associated Jacobian matrix {{${{J}}$}} which must fulfill the symplecticity condition:
\begin{equation}
{{{J}}_{ik}}~=~\frac{\partial z_{i}(s_{2})}{\partial z_{k}(s_{1})}~~~~{\mathrm{and}}~~~~{{{{{J}}}}}~~~{\mathrm{must~fulfill:}}~~~{{{{J}}^{t}}} \cdot {{S}} \cdot {{J}} = {{S}}
\end{equation}
~~~\\
However, in general
${{{{{J}}}}_{ik}}~\neq~$const and for a truncated Taylor map it can be difficult to fulfill 
this condition for all $z$.
As a consequence, the number of independent coefficients in the Taylor expansion is reduced and the
complete, symplectic Taylor map requires more coefficients than necessary \cite{chao01}.

~~~\\
The explicit map for a sextupole is:
\begin{equation}
\begin{array}{lll}
 x_{2}  &=  {\textstyle{x_{1} + L x_{1}'}}  &-~k_{2} \left(\dfrac{L^{2}}{4}(x_{1}^{2} - y_{1}^{2}) + \dfrac{L^{3}}{12}(x_{1}x_{1}' - y_{1}y_{1}') + \dfrac{L^{4}}{24}(x_{1}'^{2} - y_{1}'^{2}) \right)\\[14pt]
 x_{2}'  &=  x_{1}'  &-~k_{2} \left(\dfrac{L}{2}(x_{1}^{2} - y_{1}^{2}) + \dfrac{L^{2}}{4}(x_{1}x_{1}' - y_{1}y_{1}') + \dfrac{L^{3}}{6}(x_{1}'^{2} - y_{1}'^{2}) \right)\\[14pt]
 y_{2}  &=  y_{1} + L y_{1}'  &+~k_{2} \left(\dfrac{L^{2}}{4}x_{1}y_{1} + \dfrac{L^{3}}{12}(x_{1}y_{1}' + y_{1}x_{1}') + \dfrac{L^{4}}{24}(x_{1}'y_{1}') \right)\\[14pt]
 y_{2}'  &=  y_{1}'  &+~k_{2} \left(\dfrac{L}{2}x_{1}y_{1} + \dfrac{L^{2}}{4}(x_{1}y_{1}' + y_{1}x_{1}') + \dfrac{L^{3}}{6}(x_{1}'y_{1}') \right)\\
\end{array}
\end{equation}
Writing the explicit form of the Jacobian matrix:
\begin{equation}
{{
{{{J}}_{ik}}~=~
\left( \begin{array}{cccc}
\dfrac{\partial x_{2}}{\partial x_{1}} &~~~\dfrac{\partial x_{2}}{\partial x'_{1}} &~~~\dfrac{\partial x_{2}}{\partial y_{1}} &~~~\dfrac{\partial x_{2}}{\partial y'_{1}}\\[4mm]
\dfrac{\partial x'_{2}}{\partial x_{1}} &~~~\dfrac{\partial x'_{2}}{\partial x'_{1}} &~~~\dfrac{\partial x'_{2}}{\partial y_{1}} &~~~\dfrac{\partial x'_{2}}{\partial y'_{1}}\\[4mm]
\dfrac{\partial y_{2}}{\partial x_{1}} &~~~\dfrac{\partial y_{2}}{\partial x'_{1}} &~~~{\boldsymbol{{{\dfrac{\partial y_{2}}{\partial y_{1}}}}}} &~~~\dfrac{\partial y_{2}}{\partial y'_{1}}\\[4mm]
\dfrac{\partial y'_{2}}{\partial x_{1}} &~~~\dfrac{\partial y'_{2}}{\partial x'_{1}} &~~~\dfrac{\partial y'_{2}}{\partial y_{1}} &~~~\dfrac{\partial y'_{2}}{\partial y'_{1}}\\[4mm]
\end{array}\right)
{{~~~~~{{\rightarrow {k_{2} = 0}}}~~~~~
\left( \begin{array}{cccc}
1 &~~~~L &~~~~0 &~~~~0\\[7mm]
0 &~~~~1 &~~~~0 &~~~~0\\[7mm]
0 &~~~~0 &~~~~{\boldsymbol{1}} &~~~~L\\[7mm]
0 &~~~~0 &~~~~0 &~~~~1\\[7mm]
\end{array}\right)}}
}}
\end{equation}
\vskip 1mm
~~\\
For {{$k_{2}~\neq~0$}} the coefficients depend on initial values, e.g.:
\begin{equation}
  {\boldsymbol{{\frac{\partial y_{2}}{\partial y_{1}}}}}~=~1~+~{\underbrace{k_{2}\left(\frac{L^{2}}{4}x_{1}~+~\frac{L^{3}}{12}x_{1}'  \right)}_{\mathsf{makes~it~non-symplectic}}}~~\rightarrow~~{\mathsf{Power~series~are~not~symplectic, cannot~be~used}}
\label{eq:35}
\end{equation}
~~\\
The non-symplecticity can be recovered in the case of elements with L~=~0. It becomes
small (probably small enough) when the length is small.
~~\\
As a result, the model is approximated (non-exact) by a small amount, but the symplecticity (and therefore
the physics) is ensured. An exact model but compromised (non-symplectic) integration can 
fabricate non-existing features and conceal important underlying physics.
\vskip 3mm
~~\\
The situation is rather different in the case of single pass machines.
The long term stability (and therefore symplecticity) is not an issue
and the Taylor expansion around the closed orbit is what is really needed. 
Techniques like those described below
provide exactly this in an advanced and flexible formalism.
\subsubsection{Thick and thin lenses}
All elements in a ring have a finite length and therefore should be treated as "thick lenses".
However, in general a solution for the motion in a non-linear thick element does not exist.
It has become a standard technique to avoid using approximate integration techniques to track through
thick lenses and rather perform exact tracking through thin lenses.
This approximation is improved by breaking the thick element into  several thin elements
which is equivalent to a numerical integration.
A major advantage of this technique is that "thin lens tracking" is automatically
symplectic (see eq. \ref{eq:35}).
In this context it becomes important to understand the implied approximations and
how they influence the desired results.
We proceed by an analysis of these approximations and show how "symplectic integration"
techniques can be applied to this problem.
~~\\
We demonstrate the approximation using a quadrupole.
Although an exact solution of the motion through a quadrupole exists, 
it is a useful demonstration since it can be shown
that all concepts developed here apply also to arbitrary non-linear elements.
~~\\
Let us assume the transfer map (matrix) for a thick, linearized quadrupole of length $L$ and
strength $K$:
\begin{equation}
{{M}}_{s \rightarrow s + L}=
{{
\left( \begin{array}{cc}
\mcos{L\cdot \sqrt{K}}         &\dfrac{1}{\sqrt{K}}\cdot \msin{L\cdot {\sqrt{K}}} \\[0.3cm]
-\sqrt{K}\cdot \msin{L\cdot {\sqrt{K}}}  &\mcos{L\cdot {\sqrt{K}}} \\[0.3cm]
\end{array}\right)
}}
\end{equation}
This map is exact and can be expanded as a Taylor series for a "small" length $L$:
\begin{equation}
{{M}}_{s \rightarrow s + L}=
{{L^{0}}} \cdot \left( \begin{array}{cc}
1 &0\\[0.3cm]
0 &1\\[0.3cm]
\end{array}\right)
+ {{L^{1}}} \cdot \left( \begin{array}{cc}
0 &1\\[0.3cm]
-K &0\\[0.3cm]
\end{array}\right)
+ {{L^{2}}} \cdot \left( \begin{array}{cc}
-\frac{1}{2}{K} &0\\[0.3cm]
0 &-\frac{1}{2}{K}\\[0.3cm]
\end{array}\right) + ...
\label{eq:thl}
\end{equation}
If we keep only terms up to first order in $L$ we get:
\begin{equation}
{{M}}_{s \rightarrow s + L}=
{{L^{0}}} \cdot \left( \begin{array}{cc}
1 &0\\[0.3cm]
0 &1\\[0.3cm]
\end{array}\right)
+ {{L^{1}}} \cdot \left( \begin{array}{cc}
0 &1\\[0.3cm]
-K &0\\[0.3cm]
\end{array}\right) + {{O}}(L^{2})
\end{equation}
\begin{equation}
{{M}}_{s \rightarrow s + L}=
{{
\left( \begin{array}{cc}
1           &L  \\[0.3cm]
-K\cdot L  &1 \\[0.3cm]
\end{array}\right) + {{O}}(L^{2})
}}
\label{eq:tho1}
\end{equation}
This map is precise to order ${{O}}(L^{1})$, but since we have
det ${{M}}~\neq~$1, this truncated expansion is not symplectic.                           
\subsubsection{Symplectic matrices and symplectic integration}
However, the map (\ref{eq:tho1}) can be {\underline{made}} symplectic by adding a term ${\bf{-K^{2}L^{2}}}$.
This term is of order ${{O}}(L^{2})$, i.e. does not deteriorate the approximation because
the inaccuracy is of the same order.
\begin{equation}
{{M}}_{s \rightarrow s + L}=
{
\left( \begin{array}{cc}
1           &L  \\[0.3cm]
-K\cdot L  &1 {\bf{-KL^{2}}} \\[0.3cm]
\end{array}\right)
}
\label{eq:tho3}
\end{equation}
Following the same procedure we can compute a symplectic approximation precise
to order ${{O}}(L^{2})$ from (\ref{eq:thl}) using:
\begin{equation}
{{M}}_{s \rightarrow s + L}=
{
\left( \begin{array}{cc}
1 - \frac{1}{2}KL^{2}          &L  \\[0.3cm]
-K\cdot L  &1 - \frac{1}{2}KL^{2} \\[0.3cm]
\end{array}\right) 
~~\Rightarrow~~
\left( \begin{array}{cc}
1 - \frac{1}{2}KL^{2}          &L {\bf{- \frac{1}{4}KL^{3}}} \\[0.3cm]
-K\cdot L  &1 -\frac{1}{2}KL^{2} \\[0.3cm]
\end{array}\right)
}
\end{equation}
It can be shown that 
this "symplectification" corresponds to the approximation of a quadrupole
by a single kick in the centre between two drift spaces of length $L/2$:
\begin{equation}
\left( \begin{array}{cc}
1           &\frac{1}{2}L  \\[0.3cm]
0  &1 \\[0.3cm]
\end{array}\right)
\left( \begin{array}{cc}
1           &0  \\[0.3cm]
-K\cdot L  &1 \\[0.3cm]
\end{array}\right)
\left( \begin{array}{cc}
1           &\frac{1}{2}L  \\[0.3cm]
0  &1 \\[0.3cm]
\end{array}\right) =
\left( \begin{array}{cc}
1 - \frac{1}{2}KL^{2}          &L - \frac{1}{4}KL^{3} \\[0.3cm]
-K\cdot L  &1 -\frac{1}{2}KL^{2} \\[0.3cm]
\end{array}\right)
\end{equation}
\subsubsection{Physical meaning of the symplectification}
\begin{figure}[t]
\centering\includegraphics[height= 2.00cm,width=  6.00cm]{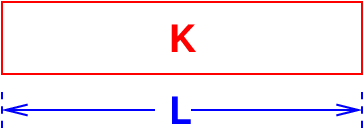}
\caption{Schematic picture of a thick quadrupole with length $L$ and strength $K$ (convention)}
\label{fig:fig4}
%\end{figure}
\vskip 2mm
%\begin{figure}[t]
\centering\includegraphics[height= 2.20cm,width=  8.20cm]{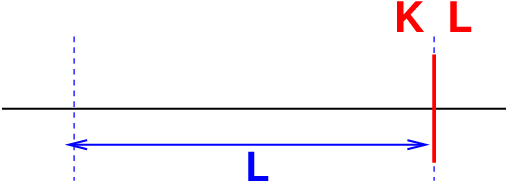}
\caption{Schematic picture of a thin quadrupole with a drift length $L$ and a kick with strength $K\cdot L$ at the {\underline{end}} of the element}
\label{fig:fig6}
\vskip 2mm
\centering\includegraphics[height= 2.20cm,width=  8.20cm]{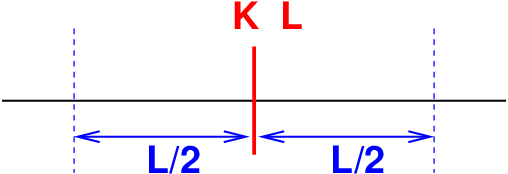}
\caption{Schematic picture of a thin quadrupole with length $L$ and a kick with strength $K\cdot L$ in the {\underline{centre}} of the element with drift space $L/2$ before and after the kick}
%\label{fig:fig6}
\label{fig:fig7}
\end{figure}

In principle, one could go on with such a scheme, but
what is a good strategy?
The main issues for using an implementation are accuracy and speed.
First, we have a look at the significance of the symplectificion,
which looks rather arbitrary.
Assume a linear element (quadrupole) of length $L$ and strength $K$ as
illustrated in Fig.\ref{fig:fig4}.
A very simple way to apply a thin-lens kick is shown in Fig.\ref{fig:fig6}.
The quadrupole is treated as a drift length of length $L$ followed by a
thin-length kick with the strength $K\cdot L$.

\vskip 3mm
~~~\\
We can compute the full matrix of the drift and the kick and obtain
\begin{eqnarray}
\left( \begin{array}{cc}
1           &0  \\[0.3cm]
-K\cdot L  &1 \\[0.3cm]
\end{array}\right)
\left( \begin{array}{cc}
1           &L  \\[0.3cm]
0  &1 \\[0.3cm]
\end{array}\right) =
\left( \begin{array}{cc}
1           &L  \\[0.3cm]
-K\cdot L  &1 -KL^{2} \\[0.3cm]
\end{array}\right).
\label{eq:19}
\end{eqnarray}
We find that this resembles the `symplectification' we applied to the truncated map of order ${\cal{O}}(L^{1})$.
Because of the symmetry, we would get the same result when we apply the kick before the drift.
~~\\
Another option is to apply the thin-lens kick in the centre of the element, as shown in Fig.\ref{fig:fig7}.
This time we apply the kick in the centre of the element, preceded and followed by drift spaces of the length
$L/2$.

%\begin{figure}[t]
%\centering\includegraphics[height= 2.20cm,width=  8.20cm]{th3.eps}
%\caption{Schematic picture of a thin quadrupole with length $L$ and a kick with strength $K\cdot L$ in the centre of the element with drift space $L/2$ before and after the kick}
%\label{fig:fig7}
%\end{figure}
~~\\
The multiplication of the three matrices leads us to
\begin{eqnarray}
{\cal{M}}_{s \rightarrow s + L}=
{{
\left( \begin{array}{cc}
1 - \frac{1}{2}KL^{2}          &L {{- \frac{1}{4}KL^{3}}} \\[0.3cm]
-K\cdot L  &1 -\frac{1}{2}KL^{2} \\[0.3cm]
\end{array}\right) + {{{\cal{O}}(L^{3})}}
}}.
\label{eq:20}
\end{eqnarray}
~~\\
This is equivalent to the truncated and symplectified map of order ${\cal{O}}(L^{2})$.
\vskip 1mm
~~\\
We can summarize the two options as follows.
\begin{itemize}[itemsep=7pt]
\item[i)]One kick at the end or entry: error (inaccuracy) is of first order ${\cal{O}}(L^{1})$.
\item[ii)]One kick in the centre: error (inaccuracy) is of second order ${\cal{O}}(L^{2})$.
\end{itemize}
We find that it is very relevant how to apply thin lenses
and the aim should be to be precise and fast (and simple to implement, for example in
a computer program).

\subsection{Symplectic integration}
We introduce now the concept of symplectic integration and start with simple examples
and generalize the concept afterwards.
Different integration methods exist which yield symplectic maps and we shall discuss
some of them which are relevant for our discussion of applications to accelerators.

\subsubsection{Can we do better than in the previous examples?}
\begin{figure}[t]
\centering\includegraphics[height= 3.20cm,width= 10.20cm]{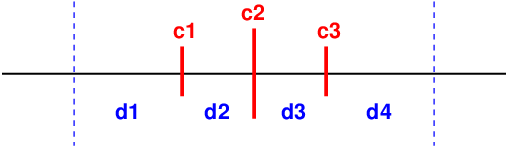}
\caption{Symplectic integrator with three thin lenses (schematic)}
\label{fig:fig8}
\end{figure}

We can try a model with three kicks~\cite{ef02}, as shown in Fig.\ref{fig:fig8}.
We assume that the kicks (c1, c2, c3) and drift spaces (d1, d2, d3, d4) are variable.
We can look for an optimization of these variables to get
the best accuracy (i.e.\ smallest deviation from the exact solution).
Since we have only thin lenses, the symplecticity is automatically ensured.

\begin{figure}[t]
\centering\includegraphics[height= 4.50cm,width= 10.20cm]{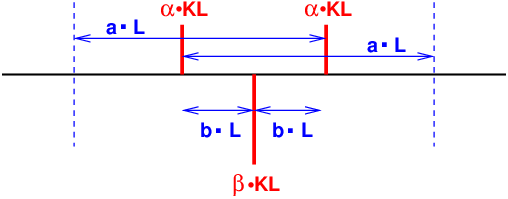}
\caption{Symplectic integrator with three thin lenses (schematic)}
\label{fig:fig9}
\end{figure}

~~\\
The results of the optimization of the strengths and the drift spaces between thin kicks are shown in Fig.\ref{fig:fig9}. The optimization of the free parameters gives the values (see Fig.\ref{fig:fig9}): 
\begin{itemize}
\item[]$a \approx 0.6756$,~~~~ $b \approx -0.1756$~~~~~~~~${({\mathsf{note:}}~2a~+~2b~~=~~1})$

\item[]$\alpha \approx 1.3512$,~~~~ $\beta \approx -1.7024$~~~~~~~~$\mathsf{(note:~2\alpha~+~\beta~~=~~1})$
\end{itemize}

~~\\
It can easily be shown \cite{ef02} that using this technique with three kicks, we obtain a ${\cal{O}}(4)$ integrator. We should expect that more kicks improve the accuracy and as an example a ${\cal{O}}(6)$ integrator would require nine kicks. This will be shown next.

\subsubsection{Yoshida formalism}
\begin{figure}[t]
\centering\includegraphics[height= 4.00cm,width= 10.20cm]{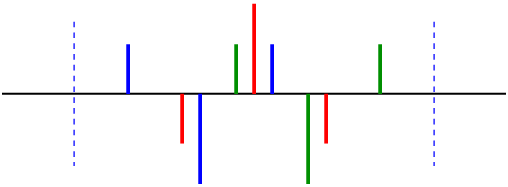}
\caption{Symplectic integrator with nine thin lenses (schematic)}
\label{fig:fig10}
\end{figure}

What we have done is a symplectic integration~\cite{ef02, yosh}. From a lower order integration scheme (one kick), we can construct a higher order scheme. Formally (for the formulation of $S_{k}(t)$ see later), we can write: from a second-order scheme (one kick) $S_{2}(t)$ we construct a fourth-order scheme
(three kicks = $3 \times 1$  kick) like
\begin{eqnarray}
S_{4}(t) = S_{2}(x_{1}t) \circ S_{2}(x_{0}t) \circ S_{2}(x_{1}t)
\label{eq:44}
\end{eqnarray}
with the coefficients
\begin{eqnarray}
 x_{0} = \frac{-2^{1/3}}{2 - 2^{1/3}} \approx -1.7024,
 \quad x_{1} = \frac{1}{2 - 2^{1/3}} \approx 1.3512.
\label{eq:45}
\end{eqnarray}
Equation (\ref{eq:44}) should be understood symbolically in that we construct
a fourth-order integrator from three second-order integrators.
As a further example from a fourth order to a sixth order, we have to use
\begin{eqnarray}
S_{6}(t) = S_{4}(x_{1}t) \circ S_{4}(x_{0}t) \circ S_{4}(x_{1}t).
\end{eqnarray}
The interpretation of this construct is that we get three times fourth order with three kicks each; we have the nine-kick, sixth-order integrator mentioned earlier. This is shown schematically in Fig.\ref{fig:fig10}. The three different colours correspond to the three fourth-order integrators, put together into a sixth-order integrator. The sixth-order integrator requires nine kicks and we have three interleaved fourth-order integrators. Going to higher order such as eighth order, we need three interleaved sixth-order integrators, altogether 27 thin lenses. The key is that this can be used as an iterative scheme to go to very high orders and can easily be implemented on a computer. More importantly, the coefficients in (\ref{eq:45}) are the same in this iterative procedure.

~~\\
As a general formula, we can express this procedure as
\begin{eqnarray}
S_{k+2}(t) = S_{k}(x_{1}t) \circ S_{k}(x_{0}t) \circ S_{k}(x_{1}t).
\end{eqnarray}
In practice, this means that each kick in the $k$th-order integrator is replaced
by a $k$th-order integrator scaled accordingly to obtain a scheme of $(k+2)$th order.

~~\\
We have used a linear map (quadrupole) to demonstrate the integration, but
it can be applied to other maps (solenoids, higher-order, non-linear maps).
It was shown~\cite{ef02, yosh} that this is indeed possible and,
most importantly, we get the same integrators.
The proof and the systematic extension can be done in the form of Lie operators.
It can be proven that this scheme gives the theoretically best possible accuracy for
a given number of thin elements.

\subsubsection{Integration of non-linear elements}
We demonstrate now this formalism for {{non-linear}} elements.
Let us assume a general case:
\begin{eqnarray}
x'' &= &f(x).
\label{eq:41}
\end{eqnarray}
The disadvantage of a kick (\ref{eq:41}) is that usually a closed solution through the
element does not exist and an integration is necessary.

The advantage of non-linear elements is that they are usually thin (at least thinner than dipoles or quadrupoles). As examples, we give the lengths of typical LHC magnets.
\begin{itemize}
\item[i)]Dipoles: $\approx$ 14.3 m.
\item[ii)]Quadrupoles: $\approx$ 2--5 m.
\item[iii)]Sextupoles, octupoles: $\approx$ 0.30 m.
\end{itemize}
We can hope to obtain a good approximation for elements with small lengths. To get an estimate, we try our simplest ${\cal{O}}(2)$ thin-lens approximation first. Therefore, we represent the non-linear element as a thin-lens kick (\ref{eq:41}) with drift spaces of $L/2$ before and after the kick.

\subsubsection{Accuracy of thin lenses: our ${\cal{O}}(2)$ model}
\begin{figure}[t]
\centering{\includegraphics*[width=50.1mm,height=25.0mm]{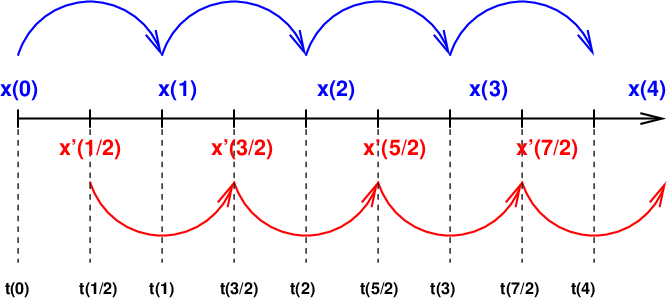}
\includegraphics*[width=34.0mm,height=30.0mm]{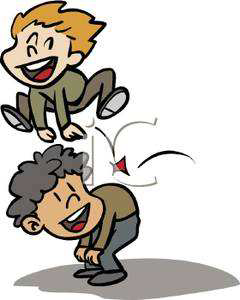}}
\caption{Schematic representation of a `leap-frog' integration scheme}
\label{fig:fig11}
\end{figure}

Assuming the general case for a non-linear kick $\Delta x'$,
\begin{eqnarray}
x'' &= &f(x)\quad(= {{\Delta x'}}),
\label{eq:21}
\end{eqnarray}
we follow the three steps:
\begin{eqnarray}
\mathrm{Step~1.}\quad\left( \begin{array}{c}
x  \\
x' \\
\end{array}\right)_{s_{1} + L/2}
=
{{
\left( \begin{array}{cc}
1 &\frac{L}{2} \\
0 &1              \\
\end{array}\right)
}}
 \circ
\left( \begin{array}{c}
x  \\
x' \\
\end{array}\right)_{s_{1}};
\end{eqnarray}

\begin{eqnarray}
\mathrm{Step~2.}\quad\left( \begin{array}{c}
x  \\
x' \\
\end{array}\right)_{s_{1} + L/2}
=
{{
\left( \begin{array}{c}
x  \\
x' + {{\Delta x'}}\\
\end{array}\right)_{s_{1} + L/2}
}};
\end{eqnarray}

\begin{eqnarray}
\mathrm{Step~3.}\quad\left( \begin{array}{c}
x  \\
x' \\
\end{array}\right)_{s_{1} + L}
=
{{
\left( \begin{array}{cc}
1 &\frac{L}{2} \\
0 &1              \\
\end{array}\right)
}}
 \circ
\left( \begin{array}{c}
x  \\
x' \\
\end{array}\right)_{s_{1} + L/2}.
\end{eqnarray}
~~\\
Using this thin-lens approximation (${\cal{O}}(2)$) gives us for the map
\begin{eqnarray}
x'(L) &\approx &x'_{0}~ +~ L~ f(x_{0}~ +~ \frac{L}{2}x'_{0}), \nonumber \\[3pt]
x(L) &\approx  &x_{0}~ +~ \frac{L}{2}~ (x'_{0}~ +~ x'(L)).
\label{eq:22}
\end{eqnarray}
Here we write $x_{0}$ and $x'_{0}$ for the coordinates and angles at the entry of the element. The result (\ref{eq:22}) corresponds to the well-known `leap-frog' algorithm/integration, schematically shown in~Fig.\ref{fig:fig11}. It is symplectic and time reversible.

~~\\
For any $x'' = f(x, x', t)$, we can solve it using
\begin{eqnarray}
x'_{i+3/2}~ &\approx~ &x'_{i+1/2}~ +~ f(x_{i+1})~ \Delta t, \nonumber \\[3pt]
x_{i+1}~ &\approx~  &x_{i}~  +~ x'_{i+1/2}~ \Delta t,
\label{eq:23}
\end{eqnarray}
which is what is known as a `leap-frog' integration (in this form).

\subsubsection{Accuracy of the `leap-frog' algorithm/integration}
The (exact) Taylor expansion gives
\begin{eqnarray}
x(L)&= &x_{0}~ +~ x'_{0}L~ +~ \frac{1}{2}f(x_{0})L^{2}~
+~~~~~ {{\frac{1}{6}}}x'_{0}f'(x_{0})L^{3}~ +~ \cdots;
\label{eq:24}
\end{eqnarray}
the (approximate) `leap-frog' algorithm (\ref{eq:23}) gives
\begin{eqnarray}
x(L)&= &x_{0}~ +~ x'_{0}L~ +~ \frac{1}{2}f(x_{0})L^{2}~ +~~~~~ {{\frac{1}{4}}}x'_{0}f'(x_{0})L^{3} +\cdots.
\label{eq:25}
\end{eqnarray}
The two methods differ from $L^{3}$ onwards and we have an integrator of order ${\cal{O}}(L^{3})$.
This had to be expected since we know from previous considerations that this type of
splitting gives such an accuracy.
For small $L$, this method is acceptable and symplectic.
~~\\
As an application, we assume a (one-dimensional) sextupole with
(definition of $k$ not unique)
\begin{eqnarray}
x'' &= &k \cdot x^{2}  =  f(x);
\label{eq:26}
\end{eqnarray}
using the thin-lens approximation using (\ref{eq:22}) gives
\begin{eqnarray}
x(L) &=  &x_{0}  + x'_{0}L + \frac{1}{2}kx^{2}_{0}L^{2}
+ \frac{1}{2}kx_{0}x'_{0}L^{3} + \frac{1}{8}kx'^{2}_{0}L^{4}, \nonumber\\[3pt]
x'(L) &= &x'_{0} + k x^{2}_{0}L + kx_{0}x'_{0}L^{2} + \frac{1}{4}kx'^{2}_{0}L^{3}.
\label{eq:27}
\end{eqnarray}
This is the map for a thick sextupole of length $L$ in thin-lens approximation, accurate to ${\cal{O}}(L^{2})$.

\section{Hamiltonian treatment of electro-magnetic fields}
A frequently asked question is why one should not just use Newton's laws and the Lorentz
force.
Some of the main reasons are:
\begin{itemize}
\item[-] Newton requires rectangular coordinates and time, trajectories with e.g. "curvature" or "torsion" need to introduce "reaction forces". (For example: LHC has locally non-planar (cork-screw) "design" orbits !).
\item[-] For linear dynamics done by ad hoc introduction of new coordinate frame.
\item[-] With Hamiltonian it is free: The formalism is "coordinate invariant", i.e. the 
equations have the same form in every coordinate system.
\item[-] The basic equations ensure that the phase space is conserved
\end{itemize}
\subsection{Lagrangian of electro-magnetic fields}
\subsubsection{Lagrangian and Hamiltonian}
It is common practice to use $q$ for the coordinates when Hamiltonian and Lagrangian
formalisms are used. This is deplorable because $q$ is also used for particle charge.
\vskip 1mm
~~\\
The motion of a particle is usually described in classical mechanics using the Langrange functional:
\begin{equation}
L(~q_{1}(t),...q_{n}(t),~~~ \dot{q_{1}}(t),...\dot{q_{n}}(t), t~)~~~~  ~{\mathsf{short:}}~~~~~L(q_{i},\dot{q_{i}},t)
\label{eq:lagr}
\end{equation}
~~\\
where $q_{1}(t),...q_{n}(t)$ are generalized coordinates and
$\dot{q_{1}}(t),...\dot{q_{n}}(t)$ the corresponding generalized velocities.
Here $q_{i}$ can stand for any coordinate and any particle, and $n$ can be a very large number. 
\vskip 2mm
~~\\
\begin{equation}
{\mathsf{The~integral~~~~~~}}
S = \int L(~q_{i}(t), \dot{q_{i}}(t), t~)~{\mathrm{d}}t 
{\mathsf{~~~~~~defines~the~action~S}}.~~~~
\label{eq:lact1}
\end{equation}
~~\\
The action $S$ is used with the Hamiltonian principle: a system moves along a path such that the action $S$ becomes stationary,
~~i.e.~ $\delta S~=~0$
~~\\
~~\\
Without proof or derivation, we quote~\cite{gold01}
 $L = T - V$,
where $T$ is the kinetic energy and $V$ the potential energy.

\begin{figure}[t]
\begin{center}
{\includegraphics*[width= 72.2mm,height=45.2mm]{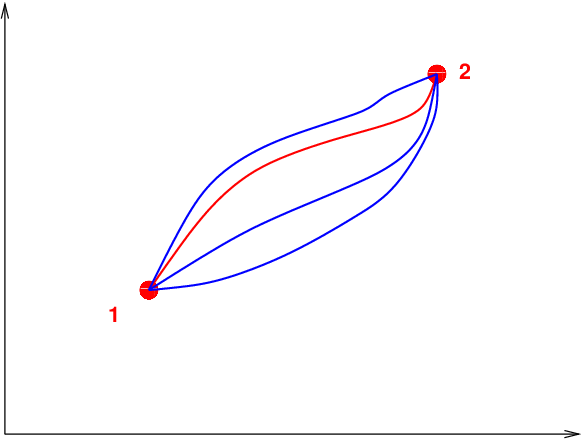}}
\end{center}
\caption{Schematic of the Hamiltonian principle. Different paths and path with minimum action}
\label{fig:fig12}
\end{figure}
~~\\
The Hamiltonian principle states that a system moves such that the action $I$ becomes an extremum when different paths are possible. This is schematically shown in Fig.\ref{fig:fig12}. The action becomes a minimum for the red path. Such concepts are known from other applications such as optics or quantum mechanics using Feynman path integrals.
~~~\\
Without proof, the action
\begin{eqnarray}
 {\mathsf{Without~proof,~the~action}}~~~~~~S = \int_{1}^{2} L(q_{i},\dot{q_{i}},t) \, {\mathrm{d}}t = \mathrm{extremum}
\end{eqnarray}
becomes an extremum (minimum) when the Lagrange function fulfils the relation~\cite{gold01}
\begin{equation}
\frac{d}{dt}\frac{\partial L}{\partial \dot{q_{i}}} - \frac{\partial L}{\partial q_{i}} = 0~~~~~~{\mathsf{(Usually~called:~~Euler~-~Lagrange~equation)}}
\label{eq:hamprinc}
\end{equation}
~~\\
It is unfortunate that the term action is used in different contexts and must not be confused with
the action-angle variables defined earlier. The action above is a functional rather than a variable.
~~\\             
~~\\
Given the Lagrangian, the Hamiltonian can be derived as:
\begin{equation}
H(\vec{q}, \vec{p}, t)~=~\sum_{i}[p_{i} \dot{q}_{i} ~-~L(\vec{q}, \vec{\dot{q}}, t)].
\label{eq:ham0}
\end{equation}
The coordinates $q_{i}$ are identical to those in the Lagrangian (\ref{eq:lagr}), 
whereas the conjugate momenta $p_{i}$
are derived from $L$ as:
\begin{equation}
p_{i}~~=~~\frac{\partial{L}}{\partial{\dot{q_{i}}}}.
\label{eq:conjp}
\end{equation}
\subsection{Hamiltonian with electro-magnetic fields}
Readers only interested in the final result can skip equations (\ref{eq:lact2}) to (\ref{eq:vel1}).
~~\\
~~\\
A key for the correct Hamiltonian is the relativistic treatment.
An intuitive derivation is presented here, a simpler and elegant
derivation should be based on 4-vectors \cite{whrel}.
The action $S$ must be a relativistic invariant and becomes (now using coordinates $x$ and velocities $v$):
\begin{equation}
S = \int L(~x_{i}(t), {v_{i}}(t), t~)~\gamma\cdot{\mathrm{d}}\tau. 
\label{eq:lact2}
\end{equation}
since the proper time $\tau$ is Lorentz invariant, and therefore also $\gamma\cdot L$.
~~\\
~~\\
The Lagrangian for a free particle is usually a function of
the velocity (see classical formula of the kinematic term),
but must not depend on its position.
~~\\
~~\\
The only Lorentz invariant with the velocity is \cite{whrel}:
\begin{equation}
U^{\mu}U_{\mu}~~=~~c^{2}~~~~~~~~{\mathsf{where}}~U~{\mathsf{(is~the~four-velocity)}}
\label{eq:invvel}
\end{equation}
~~\\
For the Lagrangian of a (relativistic) free particle we must write
~~\\
\begin{equation}
L_{free}~=~-mc^{2}\sqrt{1~-~\beta_{r}^{2}}~=~-mc^{2}\sqrt{1~-~(\frac{v}{c})^{2}}~=~-\frac{m c^{2}}{\gamma}
\label{eq:free}
\end{equation}
~~\\
Using for the electromagnetic Lagrangian a form (without derivation, any textbook):
\begin{equation}
L~=~\frac{e}{c}~{v}\cdot\vec{A}~-~e\phi
\label{eq:lint1}
\end{equation}
~~\\
Combining (\ref{eq:free}) and (\ref{eq:lint1}) we obtain the complete Lagrangian:
\begin{equation}
L~=~-\frac{mc^{2}}{\gamma}~+~\frac{e}{c}\cdot\vec{v}\cdot\vec{A}~-~e\cdot\phi
\label{eq:full}
\end{equation}
~~\\
thus the conjugate momentum is derived as:
\begin{equation}
\vec{P}~~=~~\frac{\partial{L}}{\partial{{v_{i}}}}~~=~~\vec{p}~+~\frac{e}{c} \vec{A} 
~~~~~~~~~~\left ( {\mathsf{or}}~~~~~~\vec{P}~~=~~\vec{p}~-~\frac{q}{c} \vec{A} \right)
\label{eq:can1}
\end{equation}
where $\vec{p}$ is the ordinary kinetic momentum.
~\\
~\\
A consequence is that the canonical momentum cannot be written as:
\begin{equation}
P_{x}~=~m c \gamma \beta_{x}                       
\label{eq:can2}
\end{equation}
~\\
Using the conjugate momentum the Hamiltonian takes the simple form:
\begin{equation}
H~~=~~\vec{P}\cdot\vec{v}~-~L                  
\label{eq:ham1}
\end{equation}
The Hamiltonian must be a function of the conjugate variables P and x and after a bit 
of algebra one can eliminate $\vec{v}$ using:
\begin{equation}
\vec{v}~~=~~\frac{c\vec{P}~-~e\vec{A}}{\sqrt{(\vec{P}~-~\dfrac{e\vec{A}}{c})^{2}~+~m^{2}c^{2}}}                          
\label{eq:vel1}
\end{equation}

~~\\
With (\ref{eq:full}) and (\ref{eq:vel1}) we write the
Hamiltonian for a (ultra relativistic, i.e. $\gamma \gg 1,~\beta\approx~1$) particle in an electro-magnetic field:

\begin{equation}
{{H}}(\vec{x},\vec{p}, t) = c \sqrt{(\vec{P} - e \vec{A}(\vec{x}, t) )^{2} + m^{2}c^{2}} + e \Phi(\vec{x}, t)~~
\label{eq:ham03}
\end{equation}
where $\vec{A}(\vec{x}, t)$,  $\Phi(\vec{x}, t)$ are the vector and scalar potentials.                
~~\\
~~\\
\begin{myinterlude}
~~\\
{\sffamily{
A short {\bf{interlude}}, one may want to skip to equation (\ref{eq:ham01}) 
~~\\
~~\\
The equation (\ref{eq:ham03}) is the total energy $E$ of the particle where
the difference is the potential energy $e \phi$ and the new conjugate 
momentum $\vec{P}~=~(\vec{p}~-~\frac{e}{c}\vec{A})$,~ replacing $\vec{p}$.
~~\\
~~\\
From the classical expression
\begin{equation}
E^{2}~~=~~p^{2}c^{2}~~+~~(m c^{2})^{2}
\end{equation}
one can re-write
\begin{equation}
(W~-~e\phi)^{2}~~-~~(c\vec{P}~-~e\vec{A})^{2}~~=~~(m c^{2})^{2}
\end{equation}
The expression $(m c^{2})^{2}$ is the invariant mass \cite{whrel}, i.e.
\begin{equation}
p_{\mu}p^{\mu}~~=~~(m c)^{2}
\label{eq:invm1}
\end{equation}
with the 4-vector for the momentum \cite{whrel}:
\begin{equation}
p^{\mu}~~=~~(\frac{E}{c}, \vec{p})~~=~~\left( \frac{1}{c}(W~-~e\phi),~~ \vec{P}~-~(\frac{e}{c}\vec{A}) \right)
\label{eq:invm2}
\end{equation}
}}
\end{myinterlude}
The changes are a consequence using 4-vectors in the presence of electromagnetic fields (potentials).

~~\\
An interesting consequence of (\ref{eq:can1}) is that the momentum is linked to
the fields ($\vec{A}$) and the angle $x'$ cannot easily be derived
from the total momentum and the conjugate momentum. I.e. using $(x,~x')$ as
coordinate are strictly speaking not valid in the presence of electromagnetic
fields.
~~\\
In this context using $(x,~x')$ or $(x,p_{x})$ is not equivalent. A general, 
strong statement that $(x,~x')$ "is used in accelerator physics" is at best bizarre.
~~\\
\subsection{Hamiltonian used for accelerator physics}
\vskip 2mm
In a more convenient (and useful) form, using canonical variables $x$ and $p_{x}, p_{y}$  and the design path length {{$s$}} as independent variable (bending field $B_{0}$ in y-plane) and no electric fields (for details of the derivation see  \cite{sheehy}): 
~~\\
\begin{equation}
{{H}}~=~{\underbrace{\overbrace{-(1 + \frac{x}{\rho})}^{\mathsf{due~to~t~\rightarrow~s}}}} \cdot {\overbrace{\sqrt{(1 + \delta)^{2} - p_{x}^{2} - p_{y}^{2}}}^{kinematic}} + {\underbrace{{\overbrace{{\frac{x}{\rho} + {\frac{x^{2}}{2\rho^{2}}}}}^{{\mathsf{due~to~t}~\rightarrow~s}}}}} - {\overbrace{\frac{A_{s}(x, y)}{B_{0} \rho}}^{normalized}}
\label{eq:ham01}
\end{equation}
~~\\
~~\\
where $p~=~\sqrt{E^{2}/c^{2}~-~m^{2}c^{2}}$~~total momentum, $\delta = (p - p_{0})/p_{0}$ is relative momentum deviation. 
~~\\
~~\\
$A_{s}(x, y)$ is the (normalized) longitudinal component (i.e. along {{$s$}}) of the vector potential. 
~~\\
Only transverse field and no electric fields are considered.

~~\\
After square root expansion and sorting the $A_{s}$ contributions: \\
\vskip 1mm
\begin{equation}
{{H}} = 
\overbrace{\frac{p_{x}^{2} + p_{y}^{2}}{2(1 + \delta)}}^{kinematic} -
\overbrace{\underbrace{\frac{x\delta}{\rho}}_{bending} + \underbrace{\frac{x^{2}}{2\rho^{2}}}_{focusing}}^{dipole} +
\overbrace{\frac{k_{1}}{2}(x^{2} - y^{2})}^{quadrupole} +
\overbrace{\frac{k_{2}}{6}(x^{3} - 3xy^{2})}^{sextupole}~+~~...
\label{eq:ham02}
\end{equation}
\begin{equation}
{\mathsf{using:}}~~~~~~k_{n}~=~k^{(n)}_{n} = \frac{1}{B\rho}\frac{\partial^{n}B_{y}}{\partial x^{n}}           
~~~~~~~~~\left(k^{(s)}_{n} = \frac{1}{B\rho}\frac{\partial^{n}B_{x}}{\partial x^{n}}~\right)           
\end{equation}
\vskip 1mm
~~\\
\begin{itemize}[itemsep=3pt]
\item[-]The Hamiltonian describes the motion of a particle {\underline{through an element}}
\item[-]Each element has a component in the Hamiltonian
\item[-]Basis to extend the linear to a nonlinear formalism
\end{itemize}
~~\\   
This allows to write down a short list of Hamiltonians of some individual machine elements (3D).
~\\                         
In general for multipoles of order $n$:
~~\\
\begin{equation}
 H_{n}~=~\frac{1}{1+n} {{R}}e\left[(k_{n} +{\mathrm{i}}k^{(s)}_{n})(x +{\mathrm{i}}y)^{n+1}\right]~+~\frac{p_{x}^{2}~+~p_{y}^{2}}{2(1+\delta)}
\label{eq:ham0n}
\end{equation}
~~\\
We get for some important types (normal components $k_{n}$ only):
\vskip 2mm
\begin{flalign}{{\mathsf{drift~space:}}~~H = -\sqrt{(1 + \delta)^{2}~-~p_{x}^{2}~-~p_{y}^{2}}~~\approx~~\frac{p_{x}^{2}~ +~ p_{y}^{2}}{2(1 + \delta)}}&&\end{flalign}
%\vskip 1mm
\begin{flalign}{{\mathsf{dipole}:}~~H = -\frac{-x \delta}{\rho}~ +~ \frac{x^{2}}{2\rho^{2}}~ +~ \frac{p_{x}^{2}~ +~ p_{y}^{2}}{2(1 + \delta)}}&&\end{flalign}
%\vskip 1mm
\begin{flalign}{{\mathsf{quadrupole}:}~~H = \frac{1}{2}k_{1}(x^{2} - y^{2})~ +~ \frac{p_{x}^{2}~ +~ p_{y}^{2}}{2(1 + \delta)}}&&\end{flalign}
%\vskip 1mm
\begin{flalign}{{\mathsf{sextupole}:}~~H = \frac{1}{3}k_{2}(x^{3} - {{3 x y^{2})}}~ +~ \frac{p_{x}^{2}~ +~ p_{y}^{2}}{2(1 + \delta)}}&&\end{flalign}
%\vskip 1mm
\begin{flalign}~~&{{\mathsf{octupole}:}~~H = \frac{1}{4}k_{3}(x^{4} - {{6 x^{2} y^{2}}}~ +~ y^{4})~ +~ \frac{p_{x}^{2} + p_{y}^{2}}{2(1 + \delta)}}&&\end{flalign}
\vskip 2mm
\begin{myinterlude}
~~\\
A few remarks are required after this list of Hamiltonian for particular elements.
\begin{itemize}
\item[-] Unlike seen in many introductory textbooks and lectures, a multipole
of order $n$ is {\underline{not}} required to drive a $n$th order resonance -
nothing could be more wrong !!
\item[-] In leading order perturbation theory, only elements with an even order (and larger than 2) in the Hamiltonian can produce 
an amplitude dependent tune shift and tune spread.
\end{itemize}
~~\\
\end{myinterlude}
\vskip 2mm

\subsubsection{Lie maps and transformations}
In this chapter we would like to introduce Lie algebraic tools and Lie transformations \cite{ad01, ad02, ad03}.
We use the symbol $z_{i}~=~(x_{i},p_{i})$ where $x$ and $p$ stand for canonically conjugate
position and momentum.                             
We let $f(z)$ and $g(z)$ be any function of $x, p$ and can define the
Poisson bracket for a differential operator \cite{gold01}:
\begin{equation}
[f, g] = \sum_{i=1}^{n} \left( \frac{\partial f}{\partial x_{i}}\frac{\partial g}{\partial p_{i}} - \frac{\partial f}{\partial p_{i}}\frac{\partial g}{\partial x_{i}}\right)
\end{equation}
~~\\
Assuming that the motion of a dynamic system is defined by a Hamiltonian $H$,
we can now write for the equations of motion \cite{gold01}:
\begin{flalign}
[x_{i}, H] = \frac{\partial H}{\partial p_{i}} = \frac{d x_{i}}{dt}
~~~~~~~[p_{i}, H] = -\frac{\partial H}{\partial x_{i}} = \frac{d p_{i}}{dt}
\end{flalign}
%\begin{equation}
%[p_{i}, H] = -\frac{\partial H}{\partial x_{i}} = \frac{d p_{i}}{dt}
%\end{equation}
~~\\
\begin{flalign}
{\mathsf{If}}~~H~~{\mathsf{does~not~explicitly~depend~on~time~then}}~~~~~~~[f, H] = 0
&&\end{flalign}
implies that $f$ is an invariant of the motion.
\vskip 1mm
\begin{flalign}
{\mathsf{A~Lie~operator}}~:f:~{\mathsf{~is~defined~via~the~notation}}~~~~~ :f:g = [f, g]
&&\end{flalign}
where $:f:$ is an operator acting on the function $g$.        
~~\\
\begin{flalign}
{\mathsf{Powers~can~be~defined~as:}}~~~~~~~~~  (:f:)^{2}g = :f:(:f:g) = [f,[f, g]]  ~~~~{\mathrm{etc.}}
&&\end{flalign}
\vskip 1mm
~~\\
Using this definition, one can collect a set of useful formulae for calculations.
~~\\
Some common special (very useful) cases for $f$:
~~\\
\begin{equation}
\begin{array}{ll}
:x:~=~\dfrac{\partial}{\partial p}    &~~~~~:p:~=~-\dfrac{\partial}{\partial x}\\[4mm]
:x:^{2}~=~\overbrace{:x::x:}^{\mathsf{applied~twice}}~=~\dfrac{\partial^{2}}{\partial p^{2}}  &~~~~~:p:^{2}~=~\overbrace{:p::p:}^{\mathsf{applied~twice}}~=~\dfrac{\partial^{2}}{\partial x^{2}}\\[5mm]
:xp:~=~p\dfrac{\partial}{\partial p} - x\dfrac{\partial}{\partial x}    &~~~~~:x::p:~=~:p::x:~=~-\dfrac{\partial^{2}}{\partial x \partial p}\\[4mm]
:x^{2}:~=~2x\dfrac{\partial}{\partial p}   &~~~~~:p^{2}:~=~-2p\dfrac{\partial}{\partial x}\\[5mm]
:x^{n}:~=~n\cdot x^{n-1}\dfrac{\partial}{\partial p}   &~~~~~:p^{n}:~=~-n\cdot p^{n-1}\dfrac{\partial}{\partial x}\\
\end{array}
\label{eq:use01}
\end{equation}
~~\\
~~\\
Once powers of the Lie operators are defined, they can be used to formulated
an exponential form:
\begin{equation}
 e^{:f:}  = \sum_{i=0}^{\infty} \dfrac{1}{i!}(:f:)^{i}
\end{equation}
This expression is call a "Lie transformation".
Given the Hamiltonian $H$ of an element, the generator $f$ is this Hamiltonian multiplied by
the length $L$ of the element.
~~\\
~~\\
To evaluate a simple example,
for the case ${\textstyle{H = {-p^{2}}/{2}}}$ using the exponential form and (\ref{eq:use01}):
\begin{eqnarray}
\nonumber {{e^{\textstyle{:-Lp^{2}/2:}}}}x &= &x - \frac{1}{2}L:p^{2}:x + \frac{1}{8}L^{2}(:p^{2}:)^{2}x + .. \\[0.4cm]
                 &= &x~ +~ Lp
\end{eqnarray}
\begin{eqnarray}
\nonumber {{e^{\textstyle{:-Lp^{2}/2:}}}}p &= &p - \frac{1}{2}L:p^{2}:p + ... \\[0.4cm]
                 &= &p  
\label{eq:drift01}
\end{eqnarray}
One can easily verify that for 1D and $\delta~=~0$ 
this is the transformation of a drift space of length {{L}} (if $p~\approx~x'$) as
introduced previously.        
The function $f(x,p) = -Lp^{2}/2$ is the generator of this transformation.
~~\\
~~\\
\begin{myinterlude}
~~\\
{\sffamily{
The exact Hamiltonian in two transverse dimensions and with a relative momentum deviation $\delta$ is (full Hamiltonian with $\vec{A}(\vec{x}, t)$~=~0):
\[
H = -\sqrt{(1 + \delta)^{2}~-~p_{x}^{2}~-~p_{y}^{2}}~~~~\longrightarrow~~~~f_{drift}~=~L\cdot H
\]
The {\underline{exact}} map for a drift space is now:
\begin{eqnarray*}
x^{new} &= &x + L\cdot \frac{p_{x}}{\sqrt{(1 + \delta)^{2}~-~p_{x}^{2}~-~p_{y}^{2}}}\\
p_{x}^{new} &= &p_{x}\\[5pt]
y^{new} &= &y + L\cdot \frac{p_{y}}{\sqrt{(1 + \delta)^{2}~-~p_{x}^{2}~-~p_{y}^{2}}}\\
p_{y}^{new} &= &p_{y}\\
\end{eqnarray*}
In 2D and with $\delta~\neq~0$ it is more complicated than equation (\ref{eq:drift01}).   
In practice the map can (often) be simplified to the well known form.
}}
\end{myinterlude}
~~\\
More general, acting on the phase space coordinates:
\begin{equation}
{{e^{:f:}}} (x, p)_{1} = (x, p)_{2}
\end{equation}
is the Lie transformation which describes how to go from one point to another.
~~\\
~~\\
While a Lie operator propagates variables over an infinitesimal distance, the
Lie transformation propagates over a finite distance.
~~\\
~~\\
To illustrate this technique with some simple examples, it can be shown easily, using the formulae above, that the transformation:
\begin{equation}
e^{:\textstyle{-\frac{1}{2f}}x^{2}:}
\end{equation}
corresponds to the map of a thin quadrupole with focusing length $f$, i.e.
\begin{eqnarray*}
  x_{2} &= &x_{1} \\
  p_{2} &= &p_{1} - \frac{1}{f} x_{1}\\[3pt]
\end{eqnarray*}
A transformation of the form:
\begin{equation}
e^{\textstyle{:-\frac{1}{2}L(k^{2}x^{2} + p^{2}):}}
\end{equation}
~~\\
corresponds to the map of a thick quadrupole with length $L$ and strength $k$:
\begin{eqnarray}
  x_{2} &= &x_{1} {\mathrm{\cos}}(kL) + \frac{p_{1}}{k}{\mathrm{\sin}}(kL)\\[3pt]
  p_{2} &= &-k x_{1}{\mathrm{\sin}}(kL) + p_{1} {\mathrm{\cos}}(kL)
\end{eqnarray}
~~\\
The linear map using Twiss parameters 
in Lie representation (we shall call it $:f_{2}:$ from now on) is always of the form:
\begin{equation}
e^{\textstyle{:f_{2}:}}~~~{\mathrm{with:}}~~~f_{2}(x) = -\frac{\mu}{2}(\gamma x^{2} + 2 \alpha x p + \beta p^{2})
\end{equation}
In case of a general non-linear function f(x), i.e. with a (thin lens) kick like:
\begin{eqnarray}
  x_{2} &= &x_{1} \\
  p_{2} &= &p_{1} + f(x_{1})
\label{eq:thinlie}
\end{eqnarray}
the corresponding Lie operator can be written as:
\begin{equation}
e^{\textstyle{:h:}} = e^{\textstyle{:\int_{0}^{x} f(u) {\mathrm{d}}u:}} {~~~\mathrm{or}~~~} e^{\textstyle{:F:}} {~~~\mathrm{with}~~~} F = \int_{0}^{x} f(u) {\mathrm{d}}u.
\label{eq:genlie}
\end{equation}
An important property of the Lie transformation is that
the one turn map is the exponential of the effective Hamiltonian and the circumference $C$:
\begin{equation}
{{M}}_{ring} =e^{:-C {{H}}_{eff}:}.
\end{equation}
The main advantages of Lie transformations are that the
exponential form is always symplectic and that
a formalism exists for the concatenation of transformations.
An overview of this formalism and many examples can be found in \cite{chao01}.
As for the Lie operator, one can collect a set of useful formulae.
Another neat package with useful formulae:\\
\vskip 1mm 
With $a$ constant and $f, g, h$ arbitrary functions:
\[
:a:~=~0~~~~~\longrightarrow~~~~~e^{\textstyle{:a:}}~=~1
\]
\[
:f:a~=~0~~~~~\longrightarrow~~~~~e^{\textstyle{:f:}}a~=~a
\]
\[
e^{\textstyle{:f:}}~[g, h] = [e^{\textstyle{:f:}}g,~e^{\textstyle{:f:}}h]                    
\]
\[
e^{\textstyle{:f:}}~(g\cdot h) = e^{\textstyle{:f:}}g~\cdot~ e^{\textstyle{:f:}}h                  
\]
and very important:
\[
{{M}}~g(x)~~=~~e^{\textstyle{:f:}}~g(x)~=~g(e^{\textstyle{:f:}}~x)~~~~~~~~~~{\mathsf{e.g.}}~~~~~e^{\textstyle{:f:}}~ x^{2}~=~(e^{\textstyle{:f:}}~ x)^{2}
\]
\begin{equation}
{{M}}^{-1}~g(x)~~=~~(e^{\textstyle{:f:}})^{-1}~g(x)~=~e^{\textstyle{-:f:}}~g(x)~~~~~~~~~{\mathsf{note:}}~~~\frac{1}{e^{\textstyle{:f:}}}~~\neq~~(e^{\textstyle{:f:}})^{-1}
\end{equation}
\subsubsection{Concatenation of Lie transformations}
The concatenation is very easy
when $f$ and $g$ commute (i.e. $[f,g]~=~[g,f]~=~0$) and we have:
\begin{equation}
e^{\textstyle{{:h:}}}~=~e^{\textstyle{{:f:}}}~e^{\textstyle{{:g:}}} = e^{\textstyle{:f + g:}}
\end{equation}
The generators of the transformations can just be added.
~~\\
To combine two transformations in the general case (i.e. $[f,g]~\neq~0$) 
we can use the Baker-Campbell-Hausdorff formula (BCH) which in our convention can be written as:
\begin{equation}
\begin{array}{ll}
h = f &~+~ g ~+~ \frac{1}{2}:f:g ~+~ {\textstyle{\frac{1}{12}}}:f:^{2}g ~+~ \frac{1}{12}:g:^{2}f \\[8pt]
    &~+~ \frac{1}{24}:f::g:^{2}f ~-~ \frac{1}{720}:g:^{4}f \\[8pt]
    &~-~\frac{1}{720}:f:^{4}g ~+~ \frac{1}{360}:g::f:^{3}g ~+~ ...
\end{array}
\label{eq:bch1}
\end{equation}
In many practical cases, non-linear perturbations are localized and small compared     
to the rest of the (often linear) ring, i.e.  one of $f$ or $g$ is much smaller,
e.g. ${{f}}$ corresponds to one turn, ${{g}}$ to a small, local distortion. 
~~\\
In that case we can sum up the BCH formula to first order in the perturbation $g$ and get:
\begin{eqnarray}
~e^{\textstyle{{:h:}}}~=~e^{\textstyle{{:f:}}}~e^{\textstyle{{:g:}}}~=~\mathrm{exp}\left[:f + \left( \frac{:f:}{1 - e^{-:f:}}\right) g + {{O}}(g^{2}): \right]
\label{eq:bch2}
\end{eqnarray}
When $g$ is small compared to $f$, the first order is a good approximation.

For example, we may have a full ring $e^{:f_{2}:}$ with a small (local) distortion, e.g. a multipole $e^{:g:}$ with $g = k x^{n}$ 
then the expression:
\begin{equation}
e^{\textstyle{:h:}} = e^{\textstyle{:f_{2}:}} e^{\textstyle{:k x^{n}:}},
\end{equation}
allows the evaluation of the invariant $h$ for a single multipole of order $n$ in this case.
~~\\
~~\\
In the case that $f_{2}$, $f_{3}$, $f_{4}$, are 2nd, 3rd, 4th order polynomials (Dragt-Finn factorization \cite{df01}):
\begin{equation}
e^{\textstyle{:f:}} = e^{\textstyle{:f_{2}:}} e^{\textstyle{:f_{3}:}} e^{\textstyle{:f_{4}:}},
\end{equation}
each term is symplectic and the truncation at any order does not violate symplecticity.
~~\\
One may argue that this method is clumsy when we do the analysis of a linear system.
The reader is invited to prove this by concatenating by hand a drift space and a thin
quadrupole lens.
However, the central point of this method is that the technique works whether we do 
linear or non-linear beam dynamics and provides a formal procedure.
Lie transformations are the natural extension of the linear matrix formalism to a non-linear formalism.  
There is no need to move from one method to another as required in the traditional
treatment.
~~~\\
In the case an element is described by a Hamiltonian $H$,
the Lie map of an element of length $L$ and the Hamiltonian $H$ is:
\begin{equation}
 e^{\textstyle{-L:H:}}  = \sum_{i=0}^{\infty} \frac{1}{i!}(-L:H:)^{\textstyle{i}}
\end{equation}
For example, the Hamiltonian for a thick sextupole is:
\begin{equation}
H = \frac{1}{3}k(x^{3} - 3xy^{2}) + \frac{1}{2}(p_{x}^{2} + p_{y}^{2})
\end{equation}
To find the transformation we search for:
\begin{equation}
{{e^{\textstyle{-L:H:}}}}x~~~\mathrm{and}~~~{{e^{\textstyle{-L:H:}}}}p_{x}~~~~\mathrm{i.e.~for}
\end{equation}
\begin{equation}
 {{e^{\textstyle{-L:H:}}}}x  = \sum_{i=0}^{\infty} \frac{-L^{\textstyle{i}}}{i!}(:H:)^{\textstyle{i}}x
\end{equation}
~~\\
We can compute:
\begin{equation}
 {{:H:^{\textstyle{i}}}}x~~~{\mathsf{for~each}}~i
\end{equation}
to get:
\begin{equation}
 {{:H:^{1}}}x = -p_{x},
\end{equation}
~~\\
\begin{equation}
 {{:H:^{2}}}x = -k(x^{2} - y^{2}),
\end{equation}
~~\\
\begin{equation}
 {{:H:^{3}}}x = 2k(x p_{x} - y p_{y}),
\end{equation}
~~\\
\begin{equation}
 ....
\end{equation}
Putting the terms together one obtains: 
\begin{equation}
{{e^{\textstyle{-L:H:}}}}x = x + p_{x}L - \frac{1}{2}kL^{2}(x^{2} - y^{2}) - \frac{1}{3}kL^{3}(x p_{x} - y p_{y}) + ...
\end{equation}

\subsection{Analysis techniques - Poincare surface of section}
Under normal circumstances it is not required to examine the complete time development of a particle trajectory
around the machine.
Given the experimental fact that the trajectory can be measured only at a finite number of positions around
the machine, it is only useful to sample the trajectory periodically at a fixed position.
The plot of the rate of change of the phase space variables at the beginning (or end) of each period
is the appropriate method and also known as Poincare Surface of Section \cite{poin01}.
\begin{figure}[htb]
\centering{
\includegraphics*[width= 4.0cm,clip=]{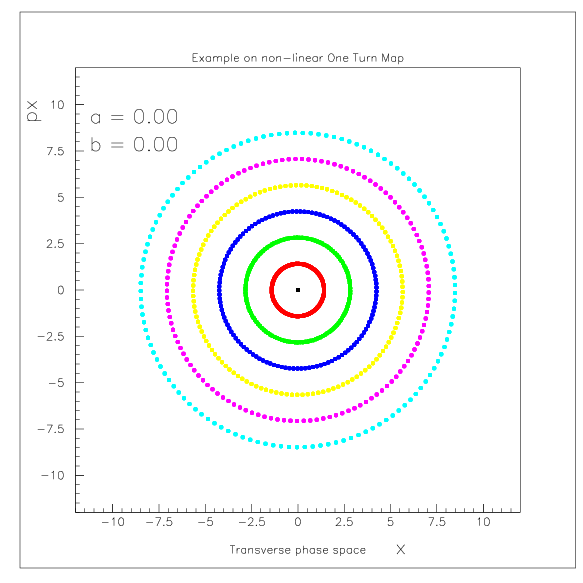}
\includegraphics*[width= 4.0cm,clip=]{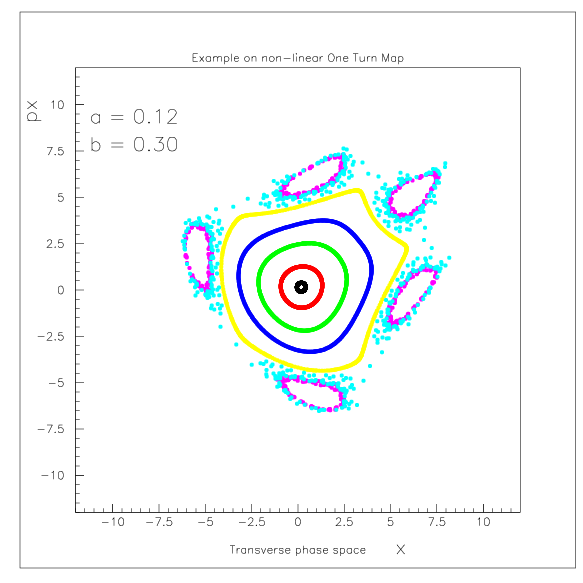}
}
\caption{Poincare surface of section of a particle near the 5th order resonances. Left without non-linear elements, right with one sextupole.}
\label{fig:04y}
\end{figure}
An example of such a plot is shown in Fig.\ref{fig:04y} where the one-dimensional phase space is plotted for a completely
linear machine (Fig.\ref{fig:04y}, left) and close to a 5th order resonance in the presence of a single non-linear
element (in this case a sextupole) in the machine (Fig.\ref{fig:04y}, right).

It shows very clearly the distortion of the phase space due to the non-linearity, the appearance of resonance islands
and chaotic behaviour between the islands.
From this plot is immediately clear that the region of stability is strongly reduced in the presence of the
non-linear element.
The main features we can observe in Fig.\ref{fig:04y} are that particles can:
\begin{itemize}
\item[$\bullet$] Move on closed curves                                                    
\item[$\bullet$] Lie on islands, i.e. jump from one island to the next from turn to turn
\item[$\bullet$] Move on chaotic trajectories                                             
\end{itemize}

The introduction of these techniques by Poincare mark a paradigm shift from the old classical treatment to
a more modern approach.
The question of long term stability of a dynamic system is not answered by getting the solution to
the differential equation of motion, but by the determination of the properties of the surface
where the motion is mapped out.
Independent how this surface of section is obtained, i.e. by analytical or numerical methods, 
its analysis is the key to understand the stability.
\subsection{Analysis techniques - Normal forms}
\label{normal-forms}
The idea behind this technique is that maps can be transformed into Normal Forms.
This tool can be used to:
\begin{itemize}
\item[$\bullet$] Study invariants of the motion and the effective Hamiltonian
\item[$\bullet$] Extract non-linear tune shifts (detuning)
\item[$\bullet$] Perform resonance analysis                      
\end{itemize}
In the following we demonstrate the use of normal forms away from resonances.
The treatment of the beam dynamics close to resonances is beyond the
scope of this review and can be found in the literature (see e.g. \cite{ef01, chao01}).
\subsubsection{Normal form transformation - linear case}
\begin{figure}[htb]
\begin{center}
\includegraphics*[width= 6.50cm,clip=]{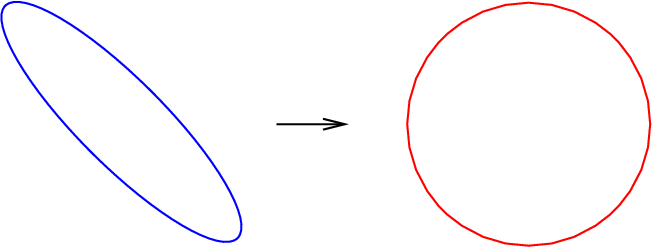}
\caption{Normal form transformation in the linear case, related to the Courant-Snyder analysis.}
\label{fig:nf01}
\end{center}
\end{figure}
The strategy is to make a transformation to get a simpler form of the
map ${M}$, e.g. a pure rotation ${{R}}(\Delta\mu)$ as schematically shown in Fig.\ref{fig:nf01}
using a transformation like:   
\begin{equation}
 {M} = {{U}} \circ {{{{R}}(\Delta\mu)} \circ {{U}}^{-1}
 ~~~~{\mathrm{or:}}~~~~{{R}}(\Delta\mu)} = {{U}}^{-1} \circ {M} \circ {{U}}
\end{equation}
with
\begin{equation}
{{U}} =
\left( \begin{array}{cc}
\sqrt{\beta(s)}  &0 \\[8pt]
-{\textstyle{\frac{\textstyle{\alpha(s)}}{\textstyle{\sqrt{\beta(s)}}}}} &\frac{1}{\textstyle{\sqrt{\beta(s)}}}  \\[8pt]
\end{array}\right)
~~~~~{\mathsf{and}}~~~~~{{R}} =
\left( \begin{array}{cc}
\cos(\Delta\mu) &\sin(\Delta\mu)  \\[8pt]
-\sin(\Delta\mu) &\cos(\Delta\mu)    \\[8pt]
\end{array}\right)
\end{equation}
~~\\
This transformation corresponds to the Courant-Snyder analysis in
the linear case and directly provides the phase advance and optical parameters.
The optical parameters emerge automatically from the normal form analysis of
the one-turn-map.
~~\\
~~\\
Although not required in the linear case, we demonstrate how this normal form transformation
is performed using the Lie formalism.
Starting from the general expression:
\begin{equation}
 {{{R}}(\Delta\mu)} = {{U}}^{-1} \circ {M} \circ {{U}}
\end{equation}
we know that a linear map ${M}$ in Lie representation is always:
\begin{equation}
e^{\textstyle{:f_{2}:}}~~~{\mathsf{with:}}~~~f_{2} = -\frac{\mu}{2}(\gamma x^{2} + 2 \alpha x p_{x} + \beta p_{x}^{2})
\end{equation}
therefore:
\begin{eqnarray}
 \nonumber{{{R}}(\Delta\mu)} &= {{U}}^{-1}~\circ~{e^{\textstyle{:f_{2}(x):}}}~\circ~{{U}}\\[3pt]
 ~~~                          &= e^{\textstyle{U^{-1}:f_{2}:U}}~=~e^{\textstyle{:U^{-1} f_{2}:}}
\end{eqnarray}
and (with $U^{-1}f_{2}$) $f_{2}$ expressed in the new variables $X, P_{x}$ it assumes the form:
~~\\
\begin{equation}
                   f_{2} = -\frac{\mu}{2}(X^{2} + P_{x}^{2})
~~~~{\mathsf{because:}}~~~~
\left( \begin{array}{c}
X  \\
P_{x}   \\
\end{array}\right)
= {{U}}^{-1} \left( \begin{array}{c}
x  \\
p_{x}   \\
\end{array}\right)
\end{equation}
~~\\
i.e. with the transformation ${{U}}^{-1}$ the rotation $:f_{2}:$ becomes a circle in the
transformed coordinates.
We transform to action and angle variables $J$ and $\Phi$, related to the
variables $X$ and $P_{x}$ through the transformations:
\begin{eqnarray}
X = \sqrt{2J\beta} \sin\Phi, ~~~~~P_{x} = \sqrt{\frac{2J}{\beta}}\cos\Phi   
\end{eqnarray}
With this transformation we get a simple representation
for the linear transfer map $f_{2}$:
\begin{eqnarray}
f_{2} = -\mu J~~~{\mathsf{and:}}~~~{{{R}}(\Delta\mu)}~=~e^{\textstyle{:-\Delta\mu J:}}
\end{eqnarray}
\subsubsection{Normal form transformation - non-linear case}
In the more general, non-linear case the transformation is more complicated and one
must expect that the rotation angle becomes amplitude dependent (see e.g. \cite{ef01}).
\begin{figure}[htb]
\includegraphics*[width= 6.50cm,height=6.50cm,angle=-90,clip=]{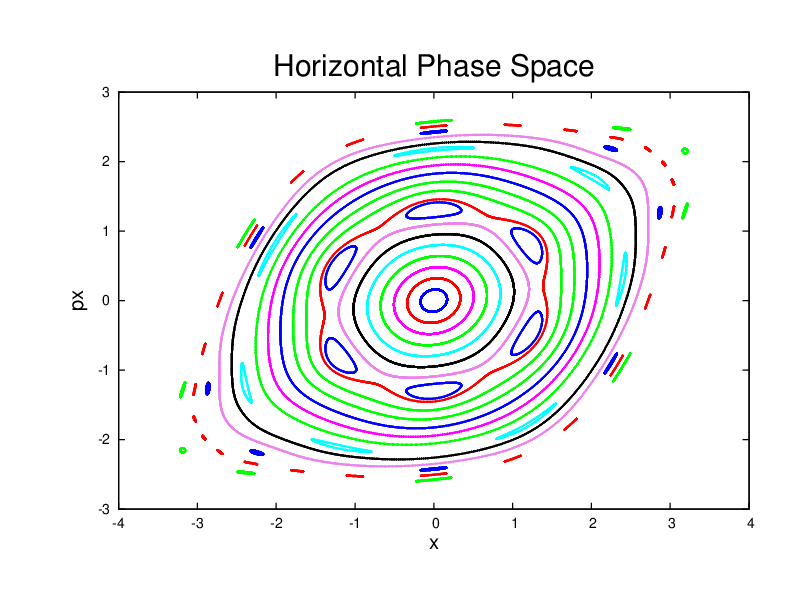}
\hskip 5mm
\includegraphics*[width= 6.50cm,height=6.50cm,angle=-90,clip=]{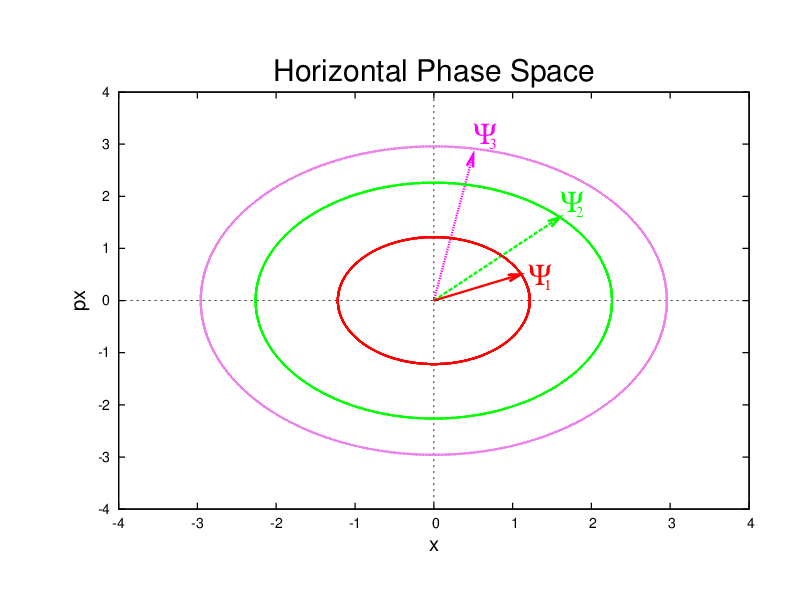}
\caption{Normal form transformation in the non-linear case, leading to amplitude dependent phase advance. The transformation was done for non-resonant amplitudes.}
\label{fig:nf02}
\end{figure}
A schematic view of this scheme is shown in Fig.\ref{fig:nf02} where the transformation leads to
the desired rotation, however the rotation frequency (phase advance) is now amplitude dependent.
~~\\
We demonstrate the power by a simple example in one dimension, but the treatment is similar
for more complex cases.
In particular, it demonstrates that this analysis using the algorithm based on Lie transforms
leads easily to the desired result.
A very detailed discussion of this method is found in \cite{ef01}.
~~\\
From the general map we have made a transformation such that the transformed map
can be expressed in the form $e^{:h_{2}:}$ where the function
$h_{2}$ is now a function only of $J_{x}, J_{y},$ and $\delta$ and it is the effective
Hamiltonian.
~~\\
~~\\
In the non-linear case and away from resonances we can get the map in a similar form:
\begin{equation}
 N = e^{\textstyle{:h_{eff}(J_{x}, J_{y},\delta):}}
\end{equation}
where the effective Hamiltonian $h_{eff}$ depends only on $J_{x}, J_{x},$ and $\delta$.
~~\\
~~\\
If the map for $h_{eff}$ corresponds to a one-turn-map, we can write for the
tunes:
\begin{equation}
Q_{x}(J_{x}, J_{y},\delta) = \frac{1}{2\pi}\frac{\partial h_{eff}}{\partial J_{x}}
\label{eq:qx}
\end{equation}
\begin{equation}
Q_{y}(J_{x}, J_{y},\delta) = \frac{1}{2\pi}\frac{\partial h_{eff}}{\partial J_{y}}
\label{eq:qy}
\end{equation}
and the change of path length:
\begin{equation}
\Delta s = -\frac{\partial h_{eff}}{\partial \delta} = \alpha_{c} \delta
\end{equation}
~~\\
In the non-linear case, particles with different $J_{x}, J_{y}, \delta$ have different tunes.
Their dependence on $J_{x}, J_{y}$ is the amplitude detuning, the dependence on $\delta$
are the chromaticities.
~~\\
The effective Hamiltonian can always be written (here to 3rd order) in a form:
\begin{eqnarray}
 h_{eff}~=~&+ &{{\mu_{x}J_{x} + \mu_{y}J_{y} + \frac{1}{2}\alpha_{c}\delta^{2}}}~~~~~~~~~~~~~~~~~~~~~~~~~~~~~~~~~~~~~~~~~~~~\\
 ~~~~~~~~~~&+ &{{c_{x1} J_{x}\delta + c_{y1} J_{y}\delta + c_{3}\delta^{3}}}~~~~~~~~~~~~~~~~~~~~~~~~~~~~~~~~~~~~~~~~~~~~~~~~~~~\\
 ~~~~~~~~~~&+ &{{c_{xx} J_{x}^{2} + c_{xy} J_{x}J_{y} + c_{yy} J_{y}^{2} + c_{x2} J_{x} \delta^{2} + c_{y2} J_{y} \delta^{2} + c_{4} \delta^{4}}}
\end{eqnarray}
and then tune depends on action {{$J$}} and momentum deviation {{$\delta$}}:
\begin{eqnarray}
Q_{x}(J_{x}, J_{y},\delta)  = \frac{1}{2\pi}\frac{\partial h_{eff}}{\partial J_{x}} = \frac{1}{2\pi}\left( {{\mu_{x}}} + \overbrace{{2c_{xx} J_{x} + c_{xy} J_{y}}}^{{detuning}} + \overbrace{{c_{x1} \delta + c_{x2} \delta^{2}}}^{{chromaticity}} \right)\\
Q_{y}(J_{x}, J_{y},\delta)  = \frac{1}{2\pi}\frac{\partial h_{eff}}{\partial J_{y}} = \frac{1}{2\pi}\left( {{\mu_{y}}} + \overbrace{{2c_{yy} J_{y} + c_{xy} J_{x}}}^{{detuning}} + \overbrace{{c_{y1} \delta + c_{y2} \delta^{2}}}^{{chromaticity}} \right)
\end{eqnarray}

The meaning of the different contributions are:
\begin{itemize}
\item[-] $\mu_{x}, \mu_{y}$: linear phase advance or 2$\pi\cdot$ i.e. the tunes for rings
\item[-] $\frac{1}{2}\alpha_{c}, c_{3}, c_{4}$: linear and nonlinear "momentum compaction"                   
\item[-] $c_{x1}, c_{y1}$: first order chromaticities                                     
\item[-] $c_{x2}, c_{y2}$: second order chromaticities                                     
\item[-] $c_{xx}, c_{xy}, c_{yy}$: detuning with amplitude                                         
\end{itemize}
\vskip 4mm
The coefficients are the various {\underline{aberrations}} of the optics.
\vskip 4mm
As a first example one can look at the effect of a single (thin) sextupole. The map is:
\begin{equation}
\begin{array}{ll}
 {\cal{{M}}} = &e^{\textstyle{-:\mu J_{x}~+~\mu J_{y}~+~\frac{1}{2}\alpha_{c}\delta^{2}:}}~{e^{\textstyle{:k (x^{3}~-~3xy^{2})}~+~\frac{p_{x}^{2} + p_{y}^{2}}{2(1+\delta)}:}}  \\
\end{array}
\end{equation}
we get for $h_{eff}$ (see e.g. {{[chao01, ef01]}}):     
\[
\begin{array}{ll}
 h_{eff} =  &\mu_{x} J_{x} + \mu_{y} J_{y} + \frac{1}{2} \alpha_{c}\delta^{2} - k D^{3}\delta^{3} - 3 k \beta_{x} J_{x} D \delta + 3 k \beta_{y} J_{y} D \delta
\end{array}
\]
Then it follows:
\begin{equation}
Q_{x}(J_{x}, J_{y},\delta) = \frac{1}{2\pi}\frac{\partial h_{eff}}{\partial J_{x}} = \frac{1}{2\pi} ({{\mu_{x}}} - {{3 k \beta_{x} D \delta}})
\end{equation}
\begin{equation}
Q_{y}(J_{x}, J_{y},\delta) = \frac{1}{2\pi}\frac{\partial h_{eff}}{\partial J_{y}} = \frac{1}{2\pi} ({{\mu_{y}}} + {{3 k \beta_{y} D \delta}})
\end{equation}
Since it was developped to first order only, there is no non-linear detuning with amplitude.
\vskip 4mm
~~\\
As a second example one can use a linear rotation followed by an octupole, the Hamiltonian is:
\begin{equation}
 {{H}} = \frac{\mu}{2}(x^{2} + p_{x}^{2}) + \delta(s - s_{0})\frac{x^{4}}{4} = \mu J + \delta(s - s_{0})\frac{x^{4}}{4}~~~~~~{\mathrm{with:}}~~J = \frac{(x^{2} + p_{x}^{2})}{2}
\label{eq:oct01}
\end{equation}
The first part of the Hamiltonian corresponds to the generator of a linear rotation and the 
second part to the localized octupole.
~~\\
The map, written in Lie representation becomes:
\begin{equation}
 M = e^{\textstyle{(-\frac{\mu}{2}:x^{2} + p_{x}^{2}:)}}~~ e^{\textstyle{:\frac{x^{4}}{4}:}} =~~ e^{\textstyle{:-\mu J:}}~ e^{\textstyle{:\frac{x^{4}}{4}:}}~ =~ R~ e^{\textstyle{:\frac{x^{4}}{4}:}}
\end{equation}
The purpose is now to find a generator $F$ for a transformation 
\begin{equation}
 e^{\textstyle{-:F:}}~ M~ e^{\textstyle{:F:}}~~=~~e^{\textstyle{-:F:}}~ e^{\textstyle{:\frac{x^{4}}{4}:}}~ e^{\textstyle{:F:}}
\end{equation}
such that the exponents of the map depend only on $J$ and not on $x$.
~~\\
~~\\
Without going through the algebra (advanced tools exist for this purpose, see e.g. \cite{ef01}) we quote the result and with
\begin{equation}
F = -\frac{1}{64}\{-5x^{4} + 3p_{x}^{4} + 6x^{2}p_{x}^{2} + x^{3}p_{x}(8\cot(\mu) + 4\cot(2\mu)) + xp_{x}^{3}(8\cot(\mu) - 4\cot(2\mu))\}
\end{equation}
~~\\
we can write the map:
\begin{equation}
 M = e^{\textstyle{-:F:}}~~e^{\textstyle{:-\mu J + {{\frac{3}{8} J^{2}}}:}}~~e^{\textstyle{:F:}}
\label{eq:oct02}
\end{equation}
the term ${{\frac{3}{8} J^{2}}}$ implies a tune shift with amplitude for an octupole.

\section[Beam dynamics with non-linearities]{Beam dynamics with non-linearities}
\label{sec-dynamics}
Following the overview of the evaluation and analysis tools, it is now 
possible to analyse and classify the behaviour of particles in the presence of 
non-linearities.
The tools presented beforehand allow a better physical insight to the mechanisms
leading to the various phenomena, the most important ones being:
\begin{itemize}
\item[$\bullet$] Amplitude detuning
\item[$\bullet$] Excitation of non-linear resonances
\item[$\bullet$] Reduction of dynamic aperture and chaotic behaviour.
\end{itemize}
This list is necessarily incomplete but will serve to demonstrate the
most important aspects.

To demonstrate these aspects, we take a realistic case and show how the 
effects emerge automatically.
\subsection{Amplitude detuning}
It was discussed in a previous section (\ref{normal-forms}) that the one-turn-map
can be transformed into a simpler map where the rotation is separated.
A consequence of the non-linearities was that the rotation frequency becomes
amplitude dependent to perform this transformation.
Therefore the amplitude detuning is directly obtained from this normal form
transformation.
\subsubsection{Amplitude detuning due to non-linearities in machine elements}
Non-linear elements cause an amplitude dependent phase advance.
The computational procedure to derive this detuning was demonstrated in the
discussion on normal for transformations in the case of an octupole 
(\ref{eq:oct01}, \ref{eq:oct02}).
This formalism is valid for any non-linear element.
~~\\
Numerous other examples can be found in \cite{ef01} and \cite{chaotig}.
\subsubsection{Amplitude detuning due to beam-beam effects}
For the demonstration we use the example of a beam-beam interaction because it
is a very complex non-linear problem and of large practical importance \cite{chao01, wh01}.

In this simplest case of one beam-beam interaction we can factorize the machine in a
linear transfer map $e^{:f_{2}:}$ and the beam-beam interaction $e^{:F:}$, i.e.:
\begin{eqnarray}
e^{\textstyle{:f_{2}:}}~\cdot~e^{\textstyle{:F:}}~=~e^{\textstyle{:{{h}}:}}
\end{eqnarray}
with
\begin{eqnarray}
f_{2}~=~ -\frac{\mu}{2} ( \frac{x^{2}}{\beta} + \beta p^{2}_{x})
\end{eqnarray}
where $\mu$ is the overall phase, i.e. the tune Q multiplied by 2$\pi$, and
$\beta$ is the $\beta$-function at the interaction point.
We assume the waist of the $\beta$-function at the collision point ($\alpha$~=~0).
The function $F(x)$ corresponds to the beam-beam potential (\ref{eq:genlie}):
\begin{eqnarray}
F(x) = \displaystyle{\int_{0}^{x}} f(u) {\mathrm{d}}u
\end{eqnarray}
For a round Gaussian beam we use for f(x) the well known expression:
\begin{eqnarray}
f(x) = \frac{2 N r_{0}}{\gamma x} ( 1 - e^{\textstyle{\dfrac{-x^{2}}{2\sigma^{2}}}})               
\end{eqnarray}
Here $N$ is the number of particles per bunch, $r_{0}$ the classical particle radius,
$\gamma$ the relativistic parameter and $\sigma$ the transverse beam size.
~~\\
For the analysis we examine the invariant $h$ which determines the one-turn-map (OTM) written
as a Lie transformation $e^{:{{h}}:}$.
The invariant $h$ is the effective Hamiltonian for this problem.
~~\\
As usual we transform to action and angle variables $J$ and $\Phi$, related to the
variables $x$ and $p_{x}$ through the transformations:
\begin{eqnarray}
x = \sqrt{2J\beta} \mathsf{sin}\Phi, ~~~~~p_{x} = \sqrt{\frac{2J}{\beta}}\cos\Phi   
\end{eqnarray}
With this transformation we get a simple representation
for the linear transfer map $f_{2}$:
\begin{eqnarray}
f_{2} = -\mu J
\end{eqnarray}
The function $F(x)$ we write as Fourier series:
\begin{eqnarray}
F(x) \Rightarrow \sum_{n=-\infty}^{\infty}  c_{n}(J) e^{\textstyle{in\Phi}}~~  {\mathsf{with}}~~ c_n(J)~=~\frac{1}{2\pi}\int_{0}^{2\pi} e^{\textstyle{-in\Phi}} F(x) {\mathrm{d}}\Phi                                       
\label{eq:cn1}
\end{eqnarray}
For the evaluation of (\ref{eq:cn1}) see \cite{chao01}.
We take some useful properties of Lie operators (e.g. \cite{ef01, chao01}):
\begin{eqnarray}
:f_{2}:g(J) = 0,~~~~~:f_{2}:e^{\textstyle{in\Phi}} = in \mu e^{\textstyle{in\Phi}},~~~~~g(:f_{2}:)e^{\textstyle{in \Phi}} = g(in \mu) e^{in\Phi}
\end{eqnarray}
and the CBH-formula for the concatenation of the maps (\ref{eq:bch2}):
\begin{eqnarray}
e^{\textstyle{:f_{2}:}}~e^{\textstyle{:F:}}~=~e^{\textstyle{:h:}}~=~\mathrm{exp}\left[:f_{2} + \left( \frac{:f_{2}:}{1 - e^{\textstyle{-:f_{2}:}}}\right) F + {{O}}(F^{2}): \right]
\end{eqnarray}
which gives immediately for $h$:
\begin{eqnarray}
h = -\mu J + \sum_{n}  c_{n}(J) \frac{i n \mu}{1 - e^{\textstyle{-i n \mu}}} e^{\textstyle{in\Phi}} ~~~~= -\mu J + \sum_{n}  c_{n}(J) \frac{n \mu}{2 \sin(\frac{n \mu}{2})} e^{\textstyle{(in\Phi + i\frac{n \mu}{2})}} 
\label{eq:h1}
\end{eqnarray}
The equation (\ref{eq:h1}) is the beam-beam perturbed invariant to first order in the perturbation using (\ref{eq:bch2}).

From (\ref{eq:h1}) we observe that for $\textstyle{\nu~= \frac{\mu}{2\pi} = \frac{p}{n}}$ resonances appear for
all integers $p$ and $n$ when $c_{n}(J) \neq 0$.

Away from resonances a normal form transformation gives: 
\begin{eqnarray}
h~=~-\mu J + c_{0}(J)~=const.
\end{eqnarray}
and the oscillating term disappears.
The first term is the linear rotation and the second term gives the amplitude dependent 
tune shift (see (\ref{eq:qx})):
\begin{eqnarray}
\Delta \mu(J) = -\frac{1}{2\pi}\frac{dc_{0}(J)}{dJ}
\end{eqnarray}
The computation of this tuneshift from the equation above can be found in the literature \cite{chao01, tp01}.
\newpage
\subsubsection{Phase space structure}
To demonstrate how this technique can be used to reconstruct the phase space structure in
the presence of non-linearities, we continue with the very non-linear problem of the
beam-beam interaction treated above.
To test our result, we compare the invariant $h$ to the
results of a particle tracking program.
~~\\
The model we use in the program is rather simple:
\begin{itemize}
\item[$\bullet$] linear transfer between interactions    
\item[$\bullet$] beam-beam kick for round beams          
\item[$\bullet$] compute action $\textstyle{J~=~\dfrac{\beta^{*}}{2 \sigma^{2}}~(\dfrac{x^{2}}{\beta^{*}}~ +~ p_{x}^{2}\beta^{*})}$  
\item[$\bullet$] compute phase $\Phi~=~{\mathsf{arctan}}(\dfrac{p_{x}}{x})$           
\item[$\bullet$] compare $J$ with $h$ as a function of the phase $\Phi$
\end{itemize}
The evaluation of the invariant (\ref{eq:h1}) is done numerically with Mathematica.
\begin{figure}[htb]
\centering{
\includegraphics*[width= 5.5cm,clip=]{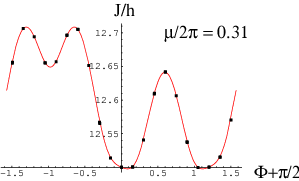}
\includegraphics*[width= 5.5cm,clip=]{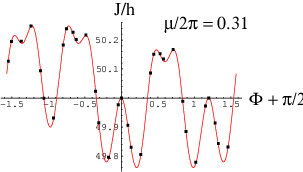}
}
\caption{Comparison: numerical and analytical model for one interaction point. Shown for 5$\sigma_{x}$ (left) and 10$\sigma_{x}$ (right). Full symbols from numerical model and solid lines from invariant (\ref{eq:h1}).}
\label{fig:04x}
\end{figure}
The comparison between the tracking results and the invariant $h$ from the analytical calculation
is shown in Fig.\ref{fig:04x} in the ($J$,$\Phi$) space.
One interaction point is used in this comparison and the particles are tracked for 1024 turns.
The symbols are the results from the tracking and the solid lines are the invariants computed as above.
The two figures are computed for amplitudes of 5~$\sigma$ and 10~$\sigma$.
The agreement between the models is excellent.
The analytic calculation was done up to the order N = 40. 
Using a lower number, the analytic model can reproduce the envelope of the tracking
results, but not the details.
The results can easily be generalized to more interaction points (\cite{wh01}).
Close to resonances these tools can reproduce the envelope of the phase space
structure (\cite{wh01}).
\subsection{Non-linear resonances}
Non-linear resonances can be excited in the presence of non-linear fields and
play a vital role for the long term stability of the particles.
\subsubsection{Resonance Condition in One Dimension}
For the special case of the beam-beam perturbed invariant (\ref{eq:h1})
we have seen that the expansion (\ref{eq:h1}) diverges when the resonance condition
for the phase advance is fulfilled, i.e.:
\begin{eqnarray}
\nu~= \dfrac{\mu}{2\pi} = \dfrac{p}{n}
\end{eqnarray}
The  formal treatment would imply to use the n-turn map with the n-turn effective Hamiltonian 
or other techniques.
This is beyond the scope of this handbook and can be found in the literature \cite{ef01, chao01}.
We should like to discuss the consequences of resonant behaviour and possible applications
in this section.
\subsubsection{Driving terms}
The treatment of the resonance map is still not fully understood and
a standard treatment using first order perturbation theory leads to a few
wrong conclusions.
In particular it is believed that a resonance cannot be excited unless
a driving term for the resonance is explicitly present in the Hamiltonian.
This implies that the related map must contain the term for a resonance in leading order
to reproduce the resonance.
This regularly leads to the conclusion that 3rd order resonances are driven by sextupoles,
4th order are driven by octupoles etc.
This is only a consequence of the perturbation theory which is often not carried
beyond leading order, and e.g. a sextupole can potentially drive resonances of any
order.
Such a treatment is valid only for special operational conditions such as
resonant extraction where strong resonant effects can be well described by a
perturbation theory.       
A detailed discussion of this misconception is given in \cite{ef01}.
A correct evaluation must be carried out to the necessary orders and the
tools presented here allow such a treatment in an easier way.
\subsection{Chromaticity and chromaticity correction}
For reasons explained earlier, sextupoles are required to correct the
chromaticities.
In large machines and in particular in colliders with insertions,
these sextupoles dominate over the non-linear effects of so-called linear elements.
\subsection{Dynamic aperture}
Often in the context of the discussion of non-linear resonance phenomena
the concept of {\it{dynamic aperture}} in introduced.
This is the maximum stable oscillation amplitude in the transverse (x,y)-space
due to non-linear fields.
It must be distinguished from the physical aperture of the vacuum chamber or other
physical restrictions such as collimators.

One of the most important tasks in the analysis of non-linear effects is to 
provide answers to the questions:
\begin{itemize}
\item[$\bullet$] Determination of the dynamic aperture
\item[$\bullet$] Maximising the dynamic aperture
\end{itemize}

The computation of the dynamic aperture is a very difficult task since no
mathematical methods are available to calculate it analytically except
for the trivial cases.
Following the concepts described earlier, the theory is much more complete
from the simulation point of view.
Therefore the standard approach to compute the dynamic aperture is done by
numerical tracking of particles.

The same techniques can be employed to maximise the dynamic aperture, in the
ideal case beyond the limits of the physical aperture.
Usually one can define tolerances for the allowed multipole components of the magnets
or the optimized parameters for colliding beams when the dominant non-linear effect
comes from beam-beam interactions.

\subsubsection{Long term stability and chaotic behaviour}
In accelerators such as particle colliders, the beams have to remain stable for
many hours and we may be asked to answer the question about stability for as
many as 10$^{9}$ turns in the machine.
This important question cannot be answered by perturbative techniques.
In the discussion of Poincare surface-of-section we have tasted the complexity
of the phase space topology and the final question is whether particles
eventually reach the entire region of the available phase space.

It was proven by Kolmogorov, Arnol'd and Moser (KAM theorem) that for weakly perturbed
systems invariant surfaces exist in the neighbourhood of integrable ones.
Poincare gave a first hint that stochastic behaviour may be generated in 
non-linear systems. In fact, higher order resonances change the topology of the
phase space and lead to the formation of island chains on an increasingly fine scale.
Satisfactory insight to the fine structure of the phase space can only be
gained with numerical computation.
Although the motion near resonances may be stochastic, the trajectories are constrained
by nearby KAM surfaces (at least in one degree of freedom) and the motion remains confined.

\subsubsection{Practical implications}
In numerical simulations where particles are tracked for millions of turns
we would like to determine the region of stability, i.e. dynamic aperture.
Since we cannot track ad infinitum, we have to specify criteria whether
a particle is stable or not.
A straightforward method is to test the particle amplitudes against
well defined apertures and declare a particle lost when the aperture
is reached.
A sufficient number of turns, usually determined by careful testing,
is required with this method.

~~\\
Usually this means to find the particle survival time as a function of
the initial amplitude.
In general the survival time decreases as the amplitude increases and 
should reach an asymptotic value at some amplitude.
The latter can be identified as the dynamic aperture.

~~\\
Other methods rely on the assumption that a particle that is unstable
in the long term, exhibits features such as a certain amount of chaotic motion.

~~\\
Typical methods to detect and quantify chaotic motion are:
\begin{itemize}
\item[$\bullet$] Frequency Map Analysis \cite{las01, las02}.
\item[$\bullet$] Lyapunov exponent \cite{ben01}.
\item[$\bullet$] Chirikov criterion \cite{chi01}.
\end{itemize}
~~\\
In all cases care must be taken to avoid numerical problems due to the 
computation techniques when a simulation over many turns is performed.

\listoffigures

\end{document}